\documentclass[12pt,twoside, a4paper]{article}
\def\pd{\partial}
\def\mc{\mathcal}

\usepackage[dvips]{graphicx}
\usepackage{amssymb}
\usepackage{amssymb,amsmath}
\usepackage{graphicx}
\input{epsf.sty}  
\begin{document}
\begin{center}
\LARGE{\textbf{Supersymmetric AdS$_2\times \Sigma_2$ solutions from
tri-sasakian truncation}}
\end{center}
\vspace{1 cm}
\begin{center}
\large{\textbf{Parinya Karndumri}}
\end{center}
\begin{center}
String Theory and Supergravity Group, Department
of Physics, Faculty of Science, Chulalongkorn University, 254 Phayathai Road, Pathumwan, Bangkok 10330, Thailand
\end{center}
E-mail: parinya.ka@hotmail.com \vspace{1 cm}\\
\begin{abstract}
A class of $AdS_2\times \Sigma_2$, with $\Sigma_2$ being
a two-sphere or a hyperbolic space, solutions within four-dimensional
$N=4$ gauged supergravity coupled to three-vector multiplets with
dyonic gauging is identified. The gauged supergravity has
non-semisimple $SO(3)\ltimes (\mathbf{T}^3,\hat{\mathbf{T}}^3)$
gauge group and can be obtained from a consistent truncation of
eleven-dimensional supergravity on a tri-sasakian manifold. The
maximally symmetric vacua contain $AdS_4$ geometries with $N=1,3$
supersymmetry corresponding to $N=1$ and $N=3$ superconformal field
theories (SCFTs) in three dimensions. We find supersymmetric
solutions of the form $AdS_2\times \Sigma_2$ preserving two supercharges.
These solutions describe twisted compactifications of the dual $N=1$
and $N=3$ SCFTs and should arise as near horizon geometries of dyonic black
holes in asymptotically $AdS_4$ space-time. Most solutions have
hyperbolic horizons although some of them exhibit spherical
horizons. These provide a new class of $AdS_2\times \Sigma_2$ geometries with
known M-theory origin.
\end{abstract}
\newpage
\section{Introduction}
Apart from giving deep insight to strongly coupled gauge theories
and string/M-theory compactifications in various dimensions, the
AdS/CFT correspondence has been recently used to explain the entropy
of asymptotically $AdS_4$ black holes
\cite{Zaffaroni_BH1,Zaffaroni_BH2,Zaffaroni_BH3}. In this context,
the black hole entropy is computed using topologically twisted index
of three-dimensional superconformal field theories (SCFTs)
compactified on a Riemann surface $\Sigma_2$
\cite{twisted_index1,twisted_index2,twisted_index3,twisted_index4,twisted_index5}.
In the dual gravity solutions, the black holes interpolate between
the asymptotically $AdS_4$ and the near horizon $AdS_2\times
\Sigma_2$ geometries. These can be interpreted as RG flows from
three dimensional SCFTs in the form of Chern-Simons-Matter (CSM)
theories possibly with flavor matters to superconformal quantum
mechanics corresponding to the $AdS_2$ geometry.
\\
\indent Along this line of research, BPS black hole solutions in
four-dimensional gauged supergravity, in particular near horizon
geometries, with known higher dimensional origins are very useful.
Most of the solutions have been studied within $N=2$ gauged
supergravities
\cite{BH_M_theory1,BH_M_theory2,AdS4_BH1,AdS4_BH2,AdS4_BH3,AdS4_BH4,AdS4_BH5},
for recent results, see \cite{Guarino_BH1,Guarino_BH2}. Many of
these solutions can be uplifted to string/M-theory since these $N=2$
gauged supergravities can be obtained either from truncations of the
maximal $N=8$ gauged supergravity, whose higher dimensional origin
is known, or direct truncations of M-theory on Sasaki-Einstein
manifolds.
\\
\indent In this work, we give an evidence for a new class of BPS
black hole solutions in the half-maximal $N=4$ gauged supergravity
with known higher dimensional origin by finding a number of new
$AdS_2\times \Sigma_2$ solutions. This gauged supergravity has
$SO(3)\ltimes (\mathbf{T}^3,\hat{\mathbf{T}}^3)$ gauge group and can
be obtained from a compactification of M-theory on a tri-sasakian
manifold \cite{N010_truncation_Cassani}. Holographic RG flows and
supersymmetric Janus solutions, describing $(1+1)$-dimensional
interfaces in the dual SCFTs have recently appeared in
\cite{trisasakian_flow}. In the present paper, we will look for
supersymmetric solutions of the form $AdS_2\times \Sigma_2$ within
this tri-sasakian compactification.
\\
\indent Apart from giving this type of solutions in gauged
supergravity with more supersymmetry, to the best of the author's
knowledge, the results are the first example of $AdS_2\times
\Sigma_2$ solutions from the truncation of M-theory on a
tri-sasakian manifold. The truncation given in
\cite{N010_truncation_Cassani} gives a reduction ansatz for
eleven-dimensional supergravity on a generic tri-sasakian manifold
including massive Kaluza-Klein modes. Among this type of manifolds,
$N^{010}$ with isometry $SU(2)\times SU(3)$ is of particular
interest. In this case, there is a non-trivial two-form giving rise
to an extra vector multiplet, see \cite{N3_spectrum1} and
\cite{N3_spectrum2} for the Kaluza-Klein spectrum of $AdS_4\times
N^{010}$. This background, discovered long ago in
\cite{Castellani_Romans}, preserves $N=3$ out of the original $N=4$
supersymmetry. There is another supersymmetric $AdS_4$ vacuum with
$SO(3)$ symmetry and $N=1$ supersymmetry, the one broken by
$AdS_4\times N^{010}$. This vacuum corresponds to $AdS_4\times
\tilde{N}^{010}$ geometry, with $\tilde{N}^{010}$ being a squashed
version of $N^{010}$.
\\
\indent Not much is known about the dual $N=1$ SCFT, but the dual
$N=3$ SCFT has been proposed in a number of previous works
\cite{Ring_N3_superfield,Shadow_N3_multiplet}, see also
\cite{Hanany_Zaffaroni1,Hanany_Zaffaroni2}. This SCFT takes the form
of a CSM theory with $SU(N)\times SU(N)$ gauge group. It is a theory
of interacting three hypermultiplets transforming in a triplet of
the $SU(3)$ flavor symmetry, and each hypermultiplet transforms as a
bifundamental under the $SU(N)\times SU(N)$ gauge group and as a
doublet of the $SU(2)_R\sim SO(3)_R$ R-symmetry. There are also a
number of previous works giving holographic studies of this theory
both in eleven-dimensional context and in the effective $N=3$ and
$N=4$ gauged supergravities
\cite{trisasakian_flow,N3_flow_Ahn1,N3_flow_Ahn2,Yi_4D_flow,N3_SU2_SU3,N3_Janus}.
Solutions given in these works are holographic RG flows, Janus
solutions and supersymmetric $AdS_2\times \Sigma_2$ solutions with magnetic
charges.
\\
\indent In this work, we consider $N=4$ gauged supergravity
constructed in the embedding tensor formalism in
\cite{N4_gauged_SUGRA}. This construction is the most general
supersymmetric gaugins of $N=4$ supergravity in which both the
``electric'' vector fields, appearing in the ungauged Lagrangian, and
their magnetic duals can participate. Therefore, magnetic and dyonic gaugings are allowed
in this formulation. Furthermore, this formulation
contains the ``purely electric'' gauged $N=4$ supergravity
constructed long time ago in \cite{Eric_N4_4D} and the non-trivial
$SL(2,\mathbb{R})$ phases of \cite{de_Roo_N4_4D,N4_Wagemans} as
special cases. We will look for supersymmetric $AdS_2\times
\Sigma_2$ solutions in the $N=4$ gauged supergravity with a dyonic
gauging of the non-semisimple group $SO(3)\ltimes
(\mathbf{T}^3,\hat{\mathbf{T}}^3)$. The solutions are required to
preserve $SO(2)\subset SO(3)_R$, so only a particular combination of
vector fields corresponding to this $SO(2)$ residual gauge symmetry
appears in the gauge covariant derivative. The strategy is essentially
similar to the wrapped brane solutions of
\cite{Maldacena_Nunez_nogo}, implementing the twist by cancelling
the spin connections on $\Sigma_2$ by the $SO(2)$ gauge connection.
\\
\indent These $AdS_2\times \Sigma_2$ solutions should appear as near
horizon geometries of supersymmetric black holes in asymptotically
$AdS_4$ space-time. Since the $N=4$ gauged supergravity admits two
supersymmetric $AdS_4$ vacua with unbroken $SO(3)_R$ symmetry and
$N=1,3$ supersymmetries, the $AdS_2\times \Sigma_2$ solutions should
be RG fixed points in one dimension of the dual $N=1,3$ SCFTs.
Although the structure of the dual $N=1$ SCFT is presently not
clear, we expect that there should be RG flows between these twisted
$N=1,3$ SCFTs on $\Sigma_2$ to one-dimensional superconformal
quantum mechanics dual to the $AdS_2\times \Sigma_2$ solutions. In
this sense, the entropy of these black holes would possibly be computed from
the topologically twisted indices of the dual $N=1,3$ SCFTs.
Furthermore, these solutions should provide a new class of $AdS_2$
geometries within M-theory.
\\
\indent The paper is organized as follow. In section \ref{N4theory},
we review $N=4$ gauged supergravity coupled to vector
multiplets and relevant formulae for uplifting the resulting solutions to eleven dimensions. The analysis of BPS equations for $SO(2)\subset SO(3)_R$ singlet scalars and Yang-Mills equations, for static black hole ansatze consistent with the symmetry of $\Sigma_2$, will be carried out in section \ref{flow_eq}. In section \ref{AdS2_solution}, we will explicitly give $AdS_2\times \Sigma_2$ solutions to the BPS flow equations. We separately consider the $N=1$ and $N=3$ cases and end the section with the uplift formulae for embedding the solutions in eleven dimensions. We finally give some conclusions and
comments on the results in section \ref{conclusions}. In the appendix, we collect the convention regarding `t Hooft matrices and give the explicit form of the Yang-Mills and BPS equations.

\section{$N=4$ gauged supergravity with dyonic gauging}\label{N4theory}
In this section, we review $N=4$ gauged supergravity in the
embedding tensor formalism following \cite{N4_gauged_SUGRA}. We
mainly focus on the bosonic Lagrangian and supersymmetry
transformations of fermions which provide basic ingredients for
finding supersymmetric solutions. Since the gauged supergravity
under consideration is known to arise from a tri-sasakian truncation
of eleven-dimensional supergravity, we will also give relevant
formulae which are useful to uplift four-dimensional solutions to
eleven dimensions. The full detail of this truncation can be found
in \cite{N010_truncation_Cassani}.

\subsection{$N=4$ gauged supergravity coupled to vector multiplets}
In the half-maximal $N=4$ supergravity in four
dimensions, the supergravity multiplet consists of the graviton
$e^{\hat{\mu}}_\mu$, four gravitini $\psi^i_\mu$, six vectors
$A_\mu^m$, four spin-$\frac{1}{2}$ fields $\chi^i$ and one complex
scalar $\tau$. The complex scalar can be parametrized by the $SL(2,\mathbb{R})/SO(2)$ coset. The supergravity
multiplet can couple to an arbitrary number $n$ of vector multiplets containing a vector
field $A_\mu$, four gaugini $\lambda^i$ and six scalars $\phi^m$.
The scalar fields can be parametrized by the $SO(6,n)/SO(6)\times
SO(n)$ coset.
\\
\indent Space-time and tangent space indices are denoted
respectively by $\mu,\nu,\ldots =0,1,2,3$ and
$\hat{\mu},\hat{\nu},\ldots=0,1,2,3$. Indices $m,n=1,\ldots, 6$ and
$i,j=1,2,3,4$ respectively describe the vector and spinor
representations of the $SO(6)_R\sim SU(4)_R$ R-symmetry or
equivalently a second-rank anti-symmetric tensor and fundamental
representations of $SU(4)_R$. The $n$ vector multiplets are labeled by
indices $a,b=1,\ldots, n$. All the fields in the vector multiplets
will accordingly carry an additional index in the form of
$(A^a_\mu,\lambda^{ia},\phi^{ma})$.
\\
\indent All fermionic fields and the supersymmetry parameters
transform in the fundamental representation of $SU(4)_R$ R-symmetry
with the chirality projections
\begin{equation}
\gamma_5\psi^i_\mu=\psi^i_\mu,\qquad \gamma_5\chi^i=-\chi^i,\qquad \gamma_5\lambda^i=\lambda^i\, .
\end{equation}
Similarly, for the fields transforming in the anti-fundamental
representation of $SU(4)_R$, we have
\begin{equation}
\gamma_5\psi_{\mu i}=-\psi_{\mu i},\qquad \gamma_5\chi_i=\chi_i,\qquad \gamma_5\lambda_i=-\lambda_i\, .
\end{equation}
\indent General gaugings of the matter-coupled $N=4$ supergravity
can be efficiently described by the embedding tensor $\Theta$ which
encodes the information about the embedding of any gauge group $G_0$
in the global or duality symmetry $SL(2,\mathbb{R})\times SO(6,n)$.
There are two components of the embedding tensor $\xi^{\alpha M}$
and $f_{\alpha MNP}$ with $\alpha=(+,-)$ and $M,N=(m,a)=1,\ldots,
n+6$ denoting fundamental representations of $SL(2,\mathbb{R})$ and
$SO(6,n)$, respectively. The electric vector fields
$A^{M+}=(A^m_\mu,A^a_\mu)$, appearing in the ungauged Lagrangian,
together with their magnetic dual $A^{M-}$ form a doublet under
$SL(2,\mathbb{R})$. These are denoted collectively by $A^{M\alpha}$.
In general, a subgroup of both $SL(2,\mathbb{R})$ and $SO(6,n)$ can
be gauged, and the magnetic vector fields can also participate in
the gauging. However, in this paper, we only consider gaugings with
only $f_{\alpha MNP}$ non-vanishing. We then set $\xi^{\alpha M}$ to
zero from now on. This also considerably simplifies many formulae
given below.
\\
\indent The full covariant derivative can be written as
\begin{equation}
D_\mu=\nabla_\mu-gA_\mu^{M\alpha}f_{\alpha M}^{\phantom{\alpha
M}NP}t_{NP}
\end{equation}
where $\nabla_\mu$ is the space-time covariant derivative including the spin connections.
$t_{MN}$ are $SO(6,n)$ generators which can be chosen as
\begin{equation}
(t_{MN})_P^{\phantom{P}Q}=2\delta^Q_{[M}\eta_{N]P},
\end{equation}
with $\eta_{MN}$ being the $SO(6,n)$ invariant tensor. The gauge
coupling constant $g$ can be absorbed in the embedding tensor
$\Theta$. The original gauging considered in \cite{Eric_N4_4D} only
involves electric vector fields and is called electric gauging. In
this case, only $f_{+MNP}$ are non-vanishing. In the following
discussions, we will consider dyonic gauging involving both electric
and magnetic vector fields. In this case, both $A^{M+}$ and $A^{M-}$
enter the Lagrangian, and $f_{\alpha MNP}$ with $\alpha=\pm$ are
non-vanishing. Consistency requires the presence of two-form fields
when magnetic vector fields are included. In the present case with
$\xi^{\alpha M}=0$, these two-forms transform as an anti-symmetric
tensor under $SO(6,n)$ and will be denoted by $B^{MN}_{\mu\nu}=B^{[MN]}_{\mu\nu}$.
The two-forms modify the gauge field strengths to
\begin{equation}
\mc{H}^{M\pm}=dA^{M\pm}-\frac{1}{2}\eta^{MQ}f_{\alpha QNP}A^{N\alpha}\wedge A^{P\pm}\pm\frac{1}{2}\eta^{MQ}f_{\mp QNP}B^{NP}\, .
\end{equation}
Note that for non-vanishing $f_{-MNP}$ the field strengths of
electric vectors $\mc{H}^{M+}$ have a contribution from the two-form
fields.
\\
\indent Before moving to the Lagrangian, we explicitly give the parametrization of the scalar coset
manifold $SL(2,\mathbb{R})/SO(2)\times SO(6,n)/SO(6)\times SO(n)$.
The first factor can be described by a coset representative
\begin{equation}
\mc{V}_\alpha=\frac{1}{\sqrt{\textrm{Im} \tau}}\left(
                                         \begin{array}{c}
                                           \tau \\
                                           1 \\
                                         \end{array}
                                       \right)\label{Valpha}
\end{equation}
or equivalently by a symmetric matrix
\begin{equation}
M_{\alpha\beta}=\textrm{Re} (\mc{V}_\alpha\mc{V}^*_\beta)=\frac{1}{\textrm{Im}
\tau}\left(
                                    \begin{array}{cc}
                                      |\tau|^2 & \textrm{Re} \tau \\
                                      \textrm{Re} \tau & 1 \\
                                    \end{array}
                                  \right).
\end{equation}
It should also be noted that $\textrm{Im}(\mc{V}_\alpha\mc{V}^*_\beta)=\epsilon_{\alpha\beta}$.
The complex scalar $\tau$ can also be written in terms of the dilaton $\phi$ and the axion $\chi$ as
\begin{equation}
\tau=\chi+ie^\phi\, .
\end{equation}
\indent For the $SO(6,n)/SO(6)\times SO(n)$ factor, we can introduce the
coset representative $\mc{V}_M^{\phantom{M}A}$ transforming by
left and right multiplications under $SO(6,n)$ and $SO(6)\times
SO(n)$, respectively. The $SO(6)\times SO(n)$ index will be split as
$A=(m,a)$ according to which the coset representative can be written as
$\mc{V}_M^{\phantom{M}A}=(\mc{V}_M^{\phantom{M}m},\mc{V}_M^{\phantom{M}a})$.
As an element of $SO(6,n)$, the matrix $\mc{V}_M^{\phantom{M}A}$ also
satisfies the relation
\begin{equation}
\eta_{MN}=-\mc{V}_M^{\phantom{M}m}\mc{V}_N^{\phantom{M}m}+\mc{V}_M^{\phantom{M}a}\mc{V}_N^{\phantom{M}a}\,
.
\end{equation}
As in the $SL(2,\mathbb{R})/SO(2)$ factor, the $SO(6,n)/SO(6)\times
SO(n)$ coset can also be parametrized in term of a symmetric matrix
defined by
\begin{equation}
M_{MN}=\mc{V}_M^{\phantom{M}m}\mc{V}_N^{\phantom{M}m}+\mc{V}_M^{\phantom{M}a}\mc{V}_N^{\phantom{M}a}\,
.
\end{equation}
\indent The bosonic Lagrangian of the $N=4$ gauged supergravity is given by
\begin{eqnarray}
e^{-1}\mc{L}&=&\frac{1}{2}R+\frac{1}{16}\mc{D}_\mu M_{MN}\mc{D}^\mu
M^{MN}-\frac{1}{4(\textrm{Im}\tau)^2}\pd_\mu \tau \pd^\mu \tau^*-V\nonumber \\
& &-\frac{1}{4}\textrm{Im}\,\tau M_{MN}\mc{H}^{M+}_{\mu\nu}\mc{H}^{N+\mu\nu}-\frac{1}{8}\textrm{Re}\,\tau e^{-1}\epsilon^{\mu\nu\rho\sigma}\eta_{MN}\mc{H}^{M+}_{\mu\nu}\mc{H}^{N+}_{\rho\sigma}\nonumber \\
& &-\frac{1}{2}e^{-1}\epsilon^{\mu\nu\rho\sigma}\left[f_{-MNP}A^{M-}_\mu A^{N+}_\nu\pd_\rho A^{P-}_\sigma +\frac{1}{4}f_{\alpha MNR}f_{\beta PQS}\eta^{RS}A^{M\alpha}_\mu A^{N+}_\nu A^{P\beta}_\rho A^{Q-}_\sigma \right.\nonumber \\
& &-\frac{1}{4}f_{-MNP}B^{NP}_{\mu\nu}\left(2\pd_\rho A^{M-}_\sigma -\frac{1}{2}\eta^{MS}f_{\alpha SQR}A^{Q\alpha}_\rho A^{R-}_\sigma\right)\nonumber \\
& &\left.-\frac{1}{16}f_{+MNR}f_{-PQS}\eta^{RS}B^{MN}_{\mu\nu}B^{PQ}_{\rho\sigma}
\right]
\end{eqnarray}
where $e$ is the vielbein determinant. The scalar potential is given
by
\begin{eqnarray}
V&=&\frac{g^2}{16}\left[f_{\alpha MNP}f_{\beta
QRS}M^{\alpha\beta}\left[\frac{1}{3}M^{MQ}M^{NR}M^{PS}+\left(\frac{2}{3}\eta^{MQ}
-M^{MQ}\right)\eta^{NR}\eta^{PS}\right]\right.\nonumber \\
& &\left.-\frac{4}{9}f_{\alpha MNP}f_{\beta
QRS}\epsilon^{\alpha\beta}M^{MNPQRS}\right]
\end{eqnarray}
where $M^{MN}$ is the inverse of $M_{MN}$, and $M^{MNPQRS}$ is
defined by
\begin{equation}
M_{MNPQRS}=\epsilon_{mnpqrs}\mc{V}_{M}^{\phantom{M}m}\mc{V}_{N}^{\phantom{M}n}
\mc{V}_{P}^{\phantom{M}p}\mc{V}_{Q}^{\phantom{M}q}\mc{V}_{R}^{\phantom{M}r}\mc{V}_{S}^{\phantom{M}s}\label{M_6}
\end{equation}
with indices raised by $\eta^{MN}$. The covariant derivative of $M_{MN}$ is defined by
\begin{equation}
\mc{D}M_{MN}=dM_{MN}+2A^{P\alpha}\eta^{QR}f_{\alpha QP(M}M_{N)R}\, .
\end{equation}
\indent It should be pointed out here that the magnetic vectors and
the two-forms do not have kinetic terms. They are auxiliary fields
whose field equations give rise to the duality relation between
two-forms and scalars and the electric-magnetic duality of $A^{M+}$
and $A^{M-}$, respectively. Together with the Yang-Mills equations
obtained from the variation with respect to $A^{M+}$, these
equations are given by
\begin{eqnarray}
\eta_{MN}*\mc{D}\mc{H}^{N-}&=&-\frac{1}{4}{f_{+MP}}^NM_{NQ}\mc{D}M^{QP},\label{YM1}\\
\eta_{MN}*\mc{D}\mc{H}^{N+}&=&\frac{1}{4}{f_{-MP}}^NM_{NQ}\mc{D}M^{QP},\label{YM2}\\
\mc{H}^{M-}&=&\textrm{Im}\, \tau M^{MN}\eta_{NP}*\mc{H}^{P+}-\textrm{Re}\, \tau\mc{H}^{M+}\label{YM3}
\end{eqnarray}
where we have used differential form language for later
computational convenience. By substituting $\mc{H}^{M-}$ from
\eqref{YM3} in \eqref{YM1}, we obtain the usual Yang-Mills equations
for $\mc{H}^{M+}$ while equation \eqref{YM2} simply gives the
relation between the Hodge dual of the three-form field strength and
the scalars due to the usual Bianchi identity of the gauge field
strengths
\begin{equation}
\mc{F}^{M\pm}=dA^{M\pm}-\frac{1}{2}\eta^{MQ}f_{\alpha QNP}A^{N\alpha}\wedge A^{P\pm}
\end{equation}
\indent In this paper, we are interested in $N=4$ gauged
supergravity coupled to three vector multiplets. The gauge group of
interest here is a non-semisimple group $SO(3)\ltimes
(\mathbf{T}^3,\hat{\mathbf{T}}^3)\subset SO(6,3)$ described by the
following components of the embedding tensor
\begin{eqnarray}
f_{+IJ,K+6}&=&-f_{+I+3,J+6,K+6}=-2\sqrt{2}\epsilon_{IJK},\qquad I,J,K=1,2,3,\nonumber \\
f_{+I+6,J+6,K+6}&=&6\sqrt{2}k\epsilon_{IJK},\qquad f_{-I,J+6,K+6}=-4\epsilon_{IJK}\, .\label{embedding_tensor}
\end{eqnarray}
The constant $k$ is related to the four-form flux along the
four-dimensional space-time, see equation \eqref{4_form_flux} below.
\\
\indent We should also remark that we follow the convention of
\cite{N010_truncation_Cassani} in all of the computations carried
out here. In particular, the $SO(6,3)$ tensor $\eta_{MN}$ is
off-diagonal
\begin{equation}
\eta_{MN}=\left(
                     \begin{array}{ccc}
                       -\mathbf{I}_3 & \mathbf{0}_3 & \mathbf{0}_3 \\
                       \mathbf{0}_3 & \mathbf{0}_3 & \mathbf{I}_3 \\
                       \mathbf{0}_3 & \mathbf{I}_3 & \mathbf{0}_3 \\
                     \end{array}
                   \right)
\end{equation}
where $\mathbf{0}_3$ and $\mathbf{I}_3$ denote $3\times 3$ zero and
identity matrices, respectively. As a result, the computation of
$M_{MNPQRS}$ in \eqref{M_6} and parts of the supersymmetry
transformations given below which involve $\mc{V}_M^{\phantom{M}m}$ and
$\mc{V}_M^{\phantom{M}a}$ must be done with the projection to the negative and
positive eigenvalues of $\eta_{MN}$, respectively. This can be achieved by using the projection matrix
\begin{equation}
P=\left(
            \begin{array}{ccc}
                       \mathbf{0}_3 & \sqrt{2}\tilde{P}_3 & \mathbf{0}_3 \\
                       -\tilde{P}_3 & \mathbf{0}_3 & \tilde{P}_3 \\
                       \tilde{P}_3 & \mathbf{0}_3 & \tilde{P}_3 \\
                     \end{array}
                   \right)
\end{equation}
where the $3\times 3$ matrix $\tilde{P}_3$ is given by
\begin{equation}
\tilde{P}_3=\frac{1}{\sqrt{2}}\left(
            \begin{array}{ccc}
                       0 & 0 & 1 \\
                       0 & 1 & 0 \\
                       1 & 0 & 0 \\
                     \end{array}
                   \right).
\end{equation}
\indent We now turn to the supersymmetry transformations of
fermionic fields. These are given by
\begin{eqnarray}
\delta\psi^i_\mu &=&2D_\mu \epsilon^i-\frac{2}{3}gA^{ij}_1\gamma_\mu
\epsilon_j+\frac{i}{4}(\mc{V}_\alpha)^*{\mc{V}_M}^{ij}\mc{H}^{M\alpha}_{\nu\rho}\gamma^{\nu\rho}\gamma_\mu\epsilon_j,\\
\delta \chi^i &=&i\epsilon^{\alpha\beta}\mc{V}_\alpha D_\mu
\mc{V}_\beta\gamma^\mu \epsilon^i-\frac{4}{3}igA_2^{ij}\epsilon_j+\frac{i}{2}\mc{V}_\alpha{\mc{V}_M}^{ij}\mc{H}^{M\alpha}_{\mu\nu}\epsilon_j,\\
\delta \lambda^i_a&=&2i\mc{V}_a^{\phantom{a}M}D_\mu
\mc{V}_M^{\phantom{M}ij}\gamma^\mu\epsilon_j+2igA_{2aj}^{\phantom{2aj}i}\epsilon^j
-\frac{1}{4}\mc{V}_\alpha\mc{V}_{Ma}\mc{H}^{M\alpha}_{\mu\nu}\gamma^{\mu\nu}\epsilon^i\,
.
\end{eqnarray}
The fermion shift matrices are defined by
\begin{eqnarray}
A_1^{ij}&=&\epsilon^{\alpha\beta}(\mc{V}_\alpha)^*\mc{V}_{kl}^{\phantom{kl}M}\mc{V}_N^{\phantom{N}ik}
\mc{V}_P^{\phantom{P}jl}f_{\beta M}^{\phantom{\beta M}NP},\nonumber
\\
A_2^{ij}&=&\epsilon^{\alpha\beta}\mc{V}_\alpha\mc{V}_{kl}^{\phantom{kl}M}\mc{V}_N^{\phantom{N}ik}
\mc{V}_P^{\phantom{P}jl}f_{\beta M}^{\phantom{\beta M}NP},\nonumber
\\
A_{2ai}^{\phantom{2ai}j}&=&\epsilon^{\alpha\beta}\mc{V}_\alpha
\mc{V}^M_{\phantom{M}a}\mc{V}^N_{\phantom{N}ik}\mc{V}_P^{\phantom{P}jk}f_{\beta
MN}^{\phantom{\beta MN}P}
\end{eqnarray}
where $\mc{V}_M^{\phantom{M}ij}$ is defined in terms of the `t Hooft
matrices $G^{ij}_m$ and $\mc{V}_M^{\phantom{M}m}$ as
\begin{equation}
\mc{V}_M^{\phantom{M}ij}=\frac{1}{2}\mc{V}_M^{\phantom{M}m}G^{ij}_m
\end{equation}
and similarly for its inverse
\begin{equation}
\mc{V}^M_{\phantom{M}ij}=-\frac{1}{2}\mc{V}_M^{\phantom{M}m}(G^{ij}_m)^*\,
.
\end{equation}
$G^{ij}_m$ satisfy the relations
\begin{equation}
G_{mij}=(G^{ij}_m)^*=\frac{1}{2}\epsilon_{ijkl}G^{kl}_m\, .
\end{equation}
The explicit form of these matrices is given in the appendix. It should also be noted that the scalar
potential can be written in terms of $A_1$ and $A_2$ tensors as
\begin{equation}
V=-\frac{1}{3}A^{ij}_1A_{1ij}+\frac{1}{9}A^{ij}_2A_{2ij}+\frac{1}{2}A_{2ai}^{\phantom{2ai}j}
A_{2a\phantom{i}j}^{\phantom{2a}i}\, .
\end{equation}
\indent With the explicit form of $\mc{V}_\alpha$ given in
\eqref{Valpha} and equation \eqref{YM3}, it is straightforward to
derive the following identities
\begin{eqnarray}
i\mc{V}_\alpha{\mc{V}_M}^{ij}\mc{H}^{M\alpha}_{\mu\nu}\gamma^{\mu\nu}&=&-(\mc{V}_-)^{-1}{\mc{V}_M}^{ij}\mc{H}^{M+}_{\mu\nu}\gamma^{\mu\nu}(1-\gamma_5),\\
i\mc{V}_\alpha{\mc{V}_M}^a\mc{H}^{M\alpha}_{\mu\nu}\gamma^{\mu\nu}&=&-(\mc{V}_-)^{-1}{\mc{V}_M}^a\mc{H}^{M+}_{\mu\nu}\gamma^{\mu\nu}(1+\gamma_5),\\
i(\mc{V}_\alpha)^*{\mc{V}_M}^{ij}\mc{H}^{M\alpha}_{\mu\nu}\gamma^{\mu\nu}\gamma_\rho
&=&(\mc{V}_-)^{-1}{\mc{V}_M}^{ij}\mc{H}^{M+}_{\mu\nu}\gamma^{\mu\nu}\gamma_\rho(1-\gamma_5).
\end{eqnarray}
In obtaining these results, we have used the following relations for
the $SO(6,n)$ coset representative \cite{Eric_N4_4D}
\begin{eqnarray}
\eta_{MN}&=&-\frac{1}{2}\epsilon_{ijkl}{\mc{V}_M}^{ij}{\mc{V}_N}^{kl}+{\mc{V}_M}^{a}{\mc{V}_N}^{a},\qquad {\mc{V}_M}^a{\mc{V}^M}_{ij}=0,\nonumber \\
{\mc{V}_M}^{ij}{\mc{V}^{M}}_{kl}&=&-\frac{1}{2}(\delta^i_k\delta^j_l-\delta^i_l\delta^j_k),\qquad {\mc{V}_M}^a{\mc{V}^M}_b=\delta^a_b\, .
\end{eqnarray}
These relations are useful in simplifying the BPS equations
resulting from the supersymmetry transformations. Note also that
these relations are slightly different from those given in
\cite{N4_gauged_SUGRA} due to a different convention on $\mc{V}_\alpha$ in term of the scalar
$\tau$. In more detail, $\mc{V}_\alpha$ used in this paper and in
\cite{N010_truncation_Cassani} satisfies $\mc{V}_+/\mc{V}_-=\tau$ while $\mc{V}_\alpha$ used in
\cite{N4_gauged_SUGRA} gives $\mc{V}_+/\mc{V}_-=\tau^*$. This results in some sign changes in the
above equations compared to those of \cite{N4_gauged_SUGRA}.

\subsection{Uplift formulae to eleven dimensions}
As mentioned above, four-dimensional $N=4$ gauged supergravity
coupled to three vector multiplets with $SO(3)\ltimes
(\mathbf{T}^3,\hat{\mathbf{T}}^3)$ gauge group has been obtained
from a truncation of eleven-dimensional supergravity on a
tri-sasakian manifold in \cite{N010_truncation_Cassani}. We will
briefly review the structure and relevant formulae focusing on the
reduction ansatz which will be useful for embedding four-dimensional
solutions. Essentially, we simply collect some formulae without
giving detailed explanations for which we refer the interested readers
to \cite{N010_truncation_Cassani}.
\\
\indent The eleven-dimensional metric can be written as
\begin{equation}
ds^2_{11}=e^{2\varphi}ds^2_4+e^{2U}ds^2(B_\textrm{QK})+g_{IJ}(\eta^I+A_1^I)(\eta^J+A_1^J)\,
.
\end{equation}
The three-dimensional internal metric $g_{IJ}$ can be written in
terms of the vielbein as
\begin{equation}
g=Q^TQ\, .
\end{equation}
Following \cite{N010_truncation_Cassani}, we will parametrize the
matrix $Q$ in term of a product of a diagonal matrix $V$ and an
$SO(3)$ matrix $O$
\begin{equation}
Q=VO,\qquad V=\textrm{diag}(e^{V_1},e^{V_2},e^{V_3})\, .\label{Q_def}
\end{equation}
The scalar $\varphi$ is chosen to be
\begin{equation}
\varphi=-\frac{1}{2}(4U+V_1+V_2+V_3)
\end{equation}
in order to obtain the Einstein frame action in four dimensions.
$B_{\textrm{QK}}$ denotes a four-dimensional quaternionic Kahler
manifold whose explicit metric is not needed in the following
discussions.
\\
\indent The ansatz for the four-form field is given by
\begin{eqnarray}
G_4&=&H_4+H_{3I}\wedge (\eta+A_1)^I+\frac{1}{2}\epsilon_{IJK}\tilde{H}_2^I\wedge(\eta+A_1)^J
\wedge (\eta+A_1)^K+4\textrm{Tr}c\,\textrm{vol}(\textrm{QK})\nonumber \\
& &H_{1IJ}\wedge (\eta+A_1)^I\wedge J^I+\frac{1}{6}\epsilon_{IJK}d\chi\wedge(\eta+A_1)^I\wedge (\eta+A_1)^J\wedge (\eta+A_1)^K\nonumber \\
& &+H_{2I}\wedge
J^I+\epsilon_{IJL}\left[(\chi+\textrm{Tr}c)\delta_{LK}-2c_{(LK)}\right](\eta+A_1)^I\wedge(\eta+A_1)^J\wedge
J^K\, .\nonumber \\
& &
\end{eqnarray}
$c_{IJ}$ is a $3\times 3$ matrix and $\textrm{Tr}
c=\delta^{IJ}c_{IJ}$. The volume form of $B_{\textrm{QK}}$,
$\textrm{vol}(\textrm{QK})$, can be written in terms of the two-forms $J^I$
as
\begin{equation}
\textrm{vol}(\textrm{QK})=\frac{1}{6}J^I\wedge J^I\, .
\end{equation}
Various forms in the above equation are defined by
\begin{eqnarray}
H_4&=&dc_3+c_{2I}\wedge F^I_2,\qquad H_{3I}=Dc_{2I}+\epsilon_{IJK}F^J_2\wedge \tilde{c}_{1K},\nonumber \\
\tilde{H}_{2I}&=&D\tilde{c}_{1I}-2c_{2I}+\chi F_{2I},\qquad H_{2I}=Dc_{1I}+2c_{2I}+c_{JI}F^J_2,\nonumber \\
H_{1IJ}&=&Dc_{IJ}+2\epsilon_{IJK}(c_{1K}+\tilde{c}_{1K})
\end{eqnarray}
with the $SO(3)$ covariant derivative
\begin{equation}
Dc_{I_1\ldots I_n}=dc_{I_1\ldots I_n}+2\sum_{l=1}^n\epsilon_{JI_lK}A^J_1\wedge c_{I_1\ldots K\ldots I_n}\, .
\end{equation}
The $SO(3)_R$ field strengths are defined by
\begin{equation}
F^I_2=dA^I_1-\epsilon_{IJK}A^J_1\wedge A^K_1\, .
\end{equation}
It is useful to note here that the $SL(2,\mathbb{R})/SO(2)$ scalars
are given by
\begin{equation}
\tau=\chi+ie^{V_1+V_2+V_3}\, .
\end{equation}
\indent Although we will not directly need the explicit form of
$ds^2(B_{\textrm{QK}})$ and $\eta^I$'s in the remaining parts of
this paper, it is useful to give some information on the $N^{010}$
tri-sasakian manifold. $N^{010}$ is a 7-manifold with $SU(2)\times
SU(3)$ isometry. The $SU(2)$ is identified with the R-symmetry of
the dual $N=3$ SCFT while $SU(3)$ is the flavor symmetry. A simple
description of $N^{010}$ can be obtained in term of a coset manifold
$SU(3)/U(1)$. With the standard Gell-Mann matrices, the $SU(3)$
generators can be chosen to be $-\frac{i}{2}\lambda_\alpha$,
$\alpha=1,\ldots, 8$. The coset and $U(1)$ generators are
accordingly identified as
\begin{equation}
K_i=-\frac{i}{2}(\lambda_1,\lambda_2,\lambda_3,\lambda_4,\lambda_5,\lambda_6,\lambda_7),\qquad
H=-\frac{i\sqrt{3}}{2}\lambda_8\, .
\end{equation}
The vielbein on $N^{010}$ can eventually be obtained from the decomposition of the Maurer-Cartan one-form
\begin{equation}
L^{-1}dL=e^iK_i+\omega H
\end{equation}
where $L$ is the coset representative for $SU(3)/U(1)$, and $\omega$ is
the corresponding $U(1)$ connection.
\\
\indent Following \cite{N010_truncation_Cassani}, we can use the tri-sasakian structures of the form
\begin{eqnarray}
\eta^I&=&\frac{1}{2}(e^1,e^2,e^7),\nonumber\\
J^I&=&\frac{1}{8}(e^4\wedge e^5-e^3\wedge e^6,-e^3\wedge e^5-e^4\wedge e^6,e^5\wedge e^6-e^3\wedge e^4).
\end{eqnarray}
From these, we find the metric on the Quaternionic-Kahler base $B_{\textrm{QK}}$ to be
\begin{equation}
ds^2(B_{\textrm{QK}})=\frac{1}{256}\left[(e^3)^2+(e^4)^2+(e^5)^2+(e^6)^2\right]
\end{equation}
with the volume form given by
\begin{equation}
\textrm{vol}(\textrm{QK})=\frac{1}{6}J^I\wedge J^I=-\frac{1}{64}e^3\wedge e^4\wedge e^5\wedge e^6\, .
\end{equation}
As mentioned before, all of the fields appearing in the reduction of \cite{N010_truncation_Cassani} are $SU(3)$ singlets.

\section{BPS flow equations}\label{flow_eq}
In this section, we perform the analysis of Yang-Mills equations
and supersymmetry transformations in order to obtain BPS equations
for the flows between $AdS_4$ vacua and possible $AdS_2\times
\Sigma_2$ geometries. We set all fermions to zero and truncate the
bosonic fields to $SO(2)\subset SO(3)_R$ singlets. This $SO(2)$ is
generated by
\begin{equation}
\hat{X}=X_{9+}+X_{6+}+X_{3-}
\end{equation}
where the gauge generators are defined by
\begin{equation}
X_{M\alpha}=-f_{\alpha MNP}t^{NP}\, .
\end{equation}
We see that a combination of the electric vectors $A^{9+}$, $A^{6+}$
and the magnetic vector $A^{3-}$ becomes the corresponding $SO(2)$
gauge field.
\\
\indent We are interested in supersymmetric solutions of the form
$AdS_2\times \Sigma_2$ with $\Sigma_2=S^2,H^2$. We will then take
the ansatz for the four-dimensional metric to be
\begin{equation}
ds^2_4=-e^{2f(r)}dt^2+dr^2+e^{2g(r)}(d\theta^2+F(\theta)^2d\phi^2)
\end{equation}
with
\begin{equation}
F(\theta)=\sin\theta\qquad \textrm{and}\qquad F(\theta)=\sinh\theta
\end{equation}
for the $S^2$ and $H^2$, respectively. We will also use the
parameter $\kappa=\pm 1$ to denote the $S^2$ and $H^2$ cases. The
functions $f(r)$, $g(r)$ and all other fields only depend on the
radial coordinate $r$ for static solutions. With the obvious
vielbein
\begin{equation}
e^{\hat{t}}=e^{f}dt,\qquad e^{\hat{r}}=dr,\qquad e^{\hat{\theta}}=e^{g}d\theta,\qquad e^{\hat{\phi}}=e^gFd\phi,
\end{equation}
it is now straightforward to compute the spin connections of the above metric
\begin{eqnarray}
\omega^{\hat{t}\hat{r}}&=&f'e^{\hat{t}},\qquad \omega^{\hat{\theta}\hat{r}}=g'e^{\hat{\theta}},\nonumber \\
\omega^{\hat{\phi}\hat{r}}&=&g'e^{\hat{\phi}},\qquad \omega^{\hat{\theta}\hat{\phi}}=\frac{F'(\theta)}{F(\theta)}e^{-g}e^{\hat{\phi}}\, .
\end{eqnarray}
In the above expressions, we have used the hat to denote ``flat''
indices while $'$ stands for the $r$-derivative with the only
exception that $F'(\theta)=\frac{dF(\theta)}{d\theta}$. The ansatz for electric and magnetic vector fields are given by
\begin{eqnarray}
A^{M+}&=&\mc{A}^M_tdt- p^MF'(\theta)d\phi, \\
A^{M-}&=&\tilde{\mc{A}}^M_tdt- e_MF'(\theta)d\phi\label{AM_minus}
\end{eqnarray}
where we have chosen the gauge such that $A^{M\alpha}_r=0$. $p^M$
and $e_M$ correspond to magnetic and electric charges, respectively.
In the present case, only $A^{M\alpha}$ with $M=3,6,9$ are relevant.
\\
\indent We finally give the explicit form of the scalar coset representative for $SO(6,3)/SO(6)\times SO(3)$. The parametrization of \cite{N010_truncation_Cassani} which is directly related to the higher dimensional origin is given by
\begin{equation}
\mc{V}=\mc{C}\mc{Q}
\end{equation}
where the matrices $\mc{Q}$ and $\mc{C}$ are defined by
\begin{equation}
\mc{Q}=\left(
                     \begin{array}{ccc}
                       \mathbf{I}_3 & \mathbf{0}_3 & \mathbf{0}_3 \\
                       \mathbf{0}_3 & e^{-2U}Q^{-1} & \mathbf{I}_3 \\
                       \mathbf{0}_3 & \mathbf{0}_3 & e^{2U}Q^T \\
                     \end{array}
                   \right),
                   \qquad
\mc{C}=\text{exp}\,\left(
                     \begin{array}{ccc}
                       \mathbf{0}_3 & \sqrt{2}c^T & \mathbf{0}_3 \\
                       \mathbf{0}_3 & \mathbf{0}_3 & \mathbf{0}_3 \\
                       \sqrt{2}c & a & \mathbf{0}_3 \\
                     \end{array}
                   \right).
\end{equation}
For $SO(2)$ invariant scalars, the $3\times 3$ matrices $c$ and $a$ are given by
\begin{equation}
c=\left(
                     \begin{array}{ccc}
                       Z_1 & Z_3 & 0 \\
                       -Z_3 & Z_1 & 0 \\
                       0 & 0 & Z_2 \\
                     \end{array}
                   \right),
                   \qquad
a=\left(
                     \begin{array}{ccc}
                       0 & \Phi & 0 \\
                       -\Phi & 0 & 0 \\
                       0 & 0 & 0 \\
                     \end{array}
                   \right)
\end{equation}
while $Q$ can be obtained from \eqref{Q_def} with $V_2=V_1$ and $O$ being
\begin{equation}
O=\textrm{exp}\left(
                     \begin{array}{ccc}
                       0 & \beta & 0 \\
                       -\beta & 0 & 0 \\
                       0 & 0 & 0 \\
                     \end{array}
                   \right).
\end{equation}
This is a generalization of the coset representative of the
$SO(3)_R$ singlet scalars used in \cite{trisasakian_flow} in which
$\Phi=\beta=Z_3=0$, $Z_1=Z_2$ and $V_1=V_2=V_3$. In the following,
we will rename the scalars $V_3\rightarrow V_2$ such that the
complex scalar $\tau$ becomes
\begin{equation}
\tau=\chi+ie^{2V_1+V_2}\, .
\end{equation}
\indent  We now give the scalar potential for $SO(2)$ singlet scalars
\begin{eqnarray}
V&=&e^{-3(4U+2V_1+V_2)}\left[e^{4(U+V_2)}(e^{4U}+2e^{4V_1})+9k^2+4\chi^2e^{4U+2V_1} \right.\nonumber\\
& &-4e^{6U+4V_1+2V_2}(6+e^{2(U-V_1)}-e^{-2(U-V_1)})+24k\chi Z_1+16\chi^2Z_1^2\nonumber \\
& &+8\chi Z_2e^{4U+2V_1}-12k\chi Z_2+(16\chi^2-24k)Z_1Z_2+32\chi Z_1^2Z_2\nonumber \\
& &+4Z_2^2e^{4U+2V_1}+4\chi^2Z_2^2+8\chi Z_1Z_2^2+16Z_1^2Z_2^2-4\chi Z_2^3 -8Z_1Z_2^3\nonumber \\
& &\left.+6kZ_2^2+Z_2^4+2e^{2V_2}\left[e^{4U}(\chi+2Z_1-Z_2)^2+2e^{4V_1}(2Z_1+Z_2)^2\right]\right].
\end{eqnarray}
The scalars $\beta$, $\Phi$ and $Z_3$ do not appear in the
potential. It can also be checked that setting $\beta=\Phi=Z_3=0$ is
a consistent truncation. In fact, $\beta$ never appears in any
equations, so we can set it to zero. On the other hand, the
Yang-Mills equations, to be given later, demand that $\Phi$ and
$Z_3$ must be constant. Since we are interested in the flow
solutions interpolating between $AdS_2\times \Sigma_2$ and $AdS_4$
vacua, and at supersymmetric $AdS_4$ critical points, both $\Phi$
and $Z_3$ vanish. We then choose $Z_3=\Phi=0$. 
\\
\indent The kinetic terms for
the remaining scalars read
\begin{eqnarray}
\mc{L}_{\textrm{kin}}&=&-6{U'}^2-2U'(2V_1'+V_2')-2V_1^{\prime 2}-V'_1V'_2\nonumber\\
& &-\frac{1}{4}\left[3V_2^{\prime 2}+e^{-2(2V_1+V_2)}{\chi'}^2+4e^{-2(2U+V_1)}Z_1^{\prime 2}+2e^{-2(2U+V_2)}Z_2^{\prime 2}\right].
\end{eqnarray}
We now redefine the scalars such that the kinetic terms are diagonal
\begin{equation}
\tilde{V}=2V_1+V_2,\qquad \tilde{U}_1=2U+V_1,\qquad \tilde{U}_2=2U+V_2
\end{equation}
in terms of which we find
\begin{equation}
\mc{L}_{\textrm{kin}}=-\frac{1}{4}\left(4\tilde{U}_1^{\prime 2}+2\tilde{U}_2^{\prime 2}+\tilde{V}^{\prime 2}+e^{-2\tilde{V}}{\chi'}^2+4e^{-2\tilde{U}_1}Z_1^{\prime 2}
+2e^{-2\tilde{U}_2}Z_2^{\prime 2} \right).
\end{equation}
These new scalars will also be useful in the analysis of the BPS equations below.
\\
\indent The above scalar potential admits two supersymmetric $AdS_4$
vacua with $N=1$ and $N=3$ supersymmetries
\cite{N010_truncation_Cassani}. At these vacua the symmetry is
enhanced from $SO(2)$ to $SO(3)$. For convenience, before carry out
the analysis of the Yang-Mills and BPS equations, we review the
$N=3$ and $N=1$ $AdS_4$ critical points in terms of the new scalars
defined above:
\begin{eqnarray}
N=3&:&\qquad \tilde{V}=\tilde{U}_1=\tilde{U}_2=\frac{1}{2}\ln k,\qquad  V_0=-12|k|^{-\frac{3}{2}},\qquad k>0, \\
N=1&:&\qquad \tilde{U}_1=\tilde{U}_2=\ln 5+\frac{1}{2}\ln \left[-\frac{k}{15}\right],\qquad  \tilde{V}=\frac{1}{2}\ln \left[-\frac{k}{15}\right],\nonumber \\
\qquad & &\qquad  V_0=-12|k|^{-\frac{3}{2}}\sqrt{\frac{3^7}{5^5}},\qquad k<0\, .
\end{eqnarray}
$V_0$ is the cosmological constant related to the $AdS_4$ radius by
\begin{equation}
L^2=-\frac{3}{V_0}\, .
\end{equation}

\subsection{The analysis of Yang-Mills equations}
We now solve the equations of motion for the gauge fields given in
\eqref{YM1}, \eqref{YM2} and \eqref{YM3}. We should emphasize that, in the reduction of
\cite{N010_truncation_Cassani}, the magnetic vectors $A^{M-}$ with
$M=4,5,6$ do not appear in the reduction ansatz. These might arise
from the reduction of the dual internal seven-dimensional metric.
Furthermore, in this reduction, the two-form fields corresponding to
these magnetic vectors do not appear. 
\\
\indent Although the present analysis involves $A^{6+}$, we will truncate out the $A^{6-}$ in order to use the reduction ansatz of \cite{N010_truncation_Cassani} to uplift the resulting solutions to eleven dimensions. This amounts to setting $e_6$ and $\tilde{\mc{A}}_t^6$ in \eqref{AM_minus} to zero. It turns out that this truncation is consistent provided that the two-form fields are properly truncated. Therefore, we will set $e_6=\tilde{\mc{A}}_t^6=0$ in the following analysis. Note also that the vanishing of $A^{6-}$ does not mean
the covariant field strength $\mc{H}^{6-}$ vanishes although the
usual gauge field strength $\mc{F}^{6-}$ vanishes. This is due to
the fact that $\mc{H}^{6-}$ gets a contribution from the two-form
fields. 
\\
\indent In order to consistently remove
$A^{6-}$, we truncate the two-form fields to only $B^{18}$ and $B^{78}$.
With the symmetry of $AdS_2\times \Sigma_2$ background and a
particular choice of tensor gauge transformations
\begin{equation}
B^{MN}\rightarrow B^{MN}+d\Xi^{MN},
\end{equation}
we will take the ansatz for the two-forms to be
\begin{equation}
B^{78}=B(r)F(\theta)d\theta \wedge d\phi,\qquad B^{18}=\tilde{B}(r)F(\theta)d\theta \wedge d\phi\, .
\end{equation}
\indent With the explicit form of the embedding tensor, we can compute the
covariant field strengths
\begin{eqnarray}
\mc{H}^{3+}&=&{\mc{A}_t^3}'dr\wedge dt+(p^3+4B)F(\theta)d\theta\wedge d\phi,\nonumber\\
\mc{H}^{6+}&=&{\mc{A}_t^6}'dr\wedge dt+(p^6-4\tilde{B})F(\theta)d\theta\wedge d\phi,\nonumber\\
\mc{H}^{9+}&=&{\mc{A}_t^9}'dr\wedge dt+p^9F(\theta)d\theta\wedge d\phi,\nonumber\\
\mc{H}^{3-}&=&\tilde{\mc{A}}^{3\prime }_tdr\wedge dt+(e_3-2\sqrt{2}\tilde{B})F(\theta)d\theta\wedge d\phi,\nonumber\\
\mc{H}^{6-}&=&-6\sqrt{2}kBF(\theta)d\theta\wedge d\phi,\nonumber\\
\mc{H}^{9-}&=&\tilde{\mc{A}}^{9\prime}_tdr\wedge dt+(e_9-2\sqrt{2}B)F(\theta)d\theta\wedge d\phi\, .
\end{eqnarray}
Note the non-vanishing covariant field strength $\mc{H}^{6-}$,  as mentioned above, due to the contribution from the two-form fields despite $A^{6-}=0$.
\\
\indent Equations arising from \eqref{YM1} and \eqref{YM2} are explicitly
given in the appendix. They can be solved by imposing the following
conditions
\begin{eqnarray}
Z_3'&=&0,\qquad \Phi'=2Z_1Z_3'-2Z_3Z_1',\nonumber \\
B'F(\theta)dr\wedge d\theta \wedge d\phi&=&\sqrt{2}e^{-4(2U+V_1)}(3k*A^{9+}+*A^{6+}-\sqrt{2}*A^{3-}),\nonumber \\
\tilde{B}'F(\theta)dr\wedge d\theta \wedge d\phi&=&4Z_1e^{-4(2U+V_1)}(3k*A^{9+}+*A^{6+}-\sqrt{2}*A^{3-}).\label{Con_YM1}
\end{eqnarray}
The first condition implies that $Z_3$ is constant. As mentioned
above, this allows to set $Z_3=0$. The second condition then
requires that $\Phi$ is constant. We can also set $\Phi=0$. Together
with $\beta=0$, we are left with only six scalars $(U,V_1,V_2,\chi,Z_1,Z_2)$ or equivalently
$(\tilde{U}_1,\tilde{U}_2,\tilde{V},\chi,Z_1,Z_2)$.
\\
\indent We move to the last two conditions in \eqref{Con_YM1}. First
of all, the $dt\wedge dr\wedge d\theta$ component gives
\begin{equation}
3kp^9+p^6-\sqrt{2}e_3=0 \label{twist1}
\end{equation}
while the $dr\wedge d\theta\wedge d\phi$ component leads to
first-order differential equations for $B$ and $\tilde{B}$
\begin{eqnarray}
B'&=&\sqrt{2}e^{-4(2U+V_1)+2g-f}(3k\mc{A}^9_t+\mc{A}^6_t-\sqrt{2}\tilde{\mc{A}}^3_t),\label{2-form-eq1}\\
\tilde{B}'&=&-4Z_1e^{-4(2U+V_1)+2g-f}(3k\mc{A}^9_t+\mc{A}^6_t-\sqrt{2}\tilde{\mc{A}}^3_t).\label{2-form-eq2}
\end{eqnarray}
\indent After solving all of the Yang-Mills equations and Bianchi
identities, we now consider the duality equation for electric and
magnetic vector fields. These equations whose explicit form is given
in the appendix lead to the relations between
$(\mc{A}^{M\prime}_t,\tilde{\mc{A}}^{M\prime}_t)$ and scalars. We
can accordingly express the former in terms of the latter. These
relations are given by
\begin{eqnarray}
\mc{A}^{3\prime}_t&=&e^{f-2g-2(U+V_1)-3V_2}\left[e^{4U+2V_2}\left[e_3+\sqrt{2}e_9Z_2-4BZ_2+\chi(p^3+4B+\sqrt{2}Z_2)\right] \right.\nonumber \\& &+Z_2^2[2(e_3+p^3\chi)+\sqrt{2}Z_2(e_9+p^9\chi)]-4Z_2B(3k-2\chi Z_2+Z_2^2)\nonumber \\
& &\left.-2\sqrt{2}\tilde{B}(e^{4U+2V_2}+2Z_2\chi+2Z_2^2)+\sqrt{2}p^6Z_2\chi\right],\label{Ap1}\\
{\mc{A}_t^6}'&=&e^{f-2g-2(2U+V_1)-3V_2}\left[(2\sqrt{2}B-e_9-p^9\chi)e^{8U+4V_2}-p^6Z_2^2\chi \right.\nonumber \\
& &-e^{4U+2V_2}Z_2[\sqrt{2}e_3-4\tilde{B}+2e_9Z_2+\chi(\sqrt{2}p^3+2p^9Z_2)]\nonumber\\
& &+4\tilde{B}Z_2^2(\chi+Z_2)-Z_2^3[\sqrt{2}(e_3+p^3\chi)+Z_2(e_9+p^9\chi)]\nonumber \\
& &\left.+4\sqrt{2}BZ_2e^{4U+2V_2}(Z_2-\chi)+2\sqrt{2}BZ_2^2(3k-2\chi Z_2+Z_2^2)\right],\\
{\mc{A}_t^9}'&=&-e^{f-2g-2(2U+V_1)-3V_2}\left[Z_2(\sqrt{2}e_3-4\tilde{B}+e_9Z_2)-2\sqrt{2}B(3k-2\chi Z_2 +Z_2^2) \right.\nonumber \\
& &\left. +\chi(p^6-4\tilde{B}+\sqrt{2}Z_2+p^9Z_2^2)\right],\\
\tilde{\mc{A}}^{3\prime}_t&=&\frac{e^{f-2g-2V_1-V_2}}{Z_2}\left[-e^{4V_1+2V_2}[\sqrt{2}e^{4U+2V_2}p^9+Z_2(p^3+4B+\sqrt{2}p^9Z_2)] \right.\nonumber \\
& &+\chi Z_2[e_3-2\sqrt{2}\tilde{B}+\sqrt{2}e_9Z_2-4BZ_2+\chi(p^3+4B+\sqrt{2}p^9Z_2)]\nonumber \\
& &\left.+\chi e^{4U+2V_2}\left[\sqrt{2}(e_9+p^9\chi)-4B\right]\right],\\
\tilde{\mc{A}}^{9\prime}_t&=&\frac{e^{f-2g-2V_1-V_2}}{Z_2^2} \left[e^{4(U+V_1+V_2)}p^9-e^{4U+2V_1}\chi(e_9-2\sqrt{2}B+p^9\chi) \right.\nonumber \\
& &-\chi Z_2[\sqrt{2}e_3-4\tilde{B}+4\sqrt{2}B(\chi-Z_2)+2e_9Z_2+\chi(\sqrt{2}p^3+2p^9Z_2)]\nonumber \\
& &\left.
+e^{4V_1+2V_2}Z_2(\sqrt{2}p^3+4\sqrt{2}B+2p^9Z_2)\right].\label{Ap6}
\end{eqnarray}
It turns out that only $\mc{A}_t^{9}$, $\mc{A}_t^{6}$ and
$\tilde{\mc{A}}_t^{3}$ appear in other equations while the remaining
ones only appear through their derivatives. Therefore, these fields
can be integrated out.

\subsection{BPS equations for $SO(2)$ invariant scalars}
We now use the ansatz for all the fields given in the previous
section to set up the BPS equations for finding supersymmetric
solutions. We will use Majorana representation for the gamma
matrices in which all $\gamma_\mu$ are real, and
\begin{equation}
\gamma_5=i\gamma_{\hat{0}}\gamma_{\hat{r}}\gamma_{\hat{\theta}}\gamma_{\hat{\phi}}
\end{equation}
is purely imaginary. We then have, for example,
\begin{equation}
\epsilon^i=\frac{1}{2}(1+\gamma_5)\epsilon^i_M,\qquad \epsilon_i=\frac{1}{2}(1-\gamma_5)\epsilon^i_M
\end{equation}
with $\epsilon^i_M$ being four-component Majorana spinors. It follows that $\epsilon_i=(\epsilon^i)^*$.
\\
\indent We first consider the gravitino transformations. As in other
holographic solutions involving twisted compactifications of the dual
SCFTs, the strategy is to use the gauge connection to cancel the
spin connection on $\Sigma_2$. Equations from
$\delta\psi^i_{\hat{\theta}}=0$ and $\delta\psi^i_{\hat{\phi}}=0$
then reduce to the same equation. The gauge connection enters the
covariant derivative of $\epsilon^i$ through the composite
connection ${Q_j}^i$. With the $SO(2)$ singlet scalars, we find that
${Q_j}^i$ takes the form of
\begin{equation}
{Q_j}^i=\frac{1}{2}\hat{A}\left(
                     \begin{array}{cccc}
                       0 & 1 & 0 & 0 \\
                       -1 & 0 & 0 & 0 \\
                       0 & 0 & 0  & 0\\
                        0 & 0 & 0  & 0\\
                     \end{array}
                   \right)
\end{equation}
where $\hat{A}$ is given by
\begin{equation}
\hat{A}=\sqrt{2}e^{-2(2U+V_1)}(3kA^{9+}+A^{6+}-\sqrt{2}A^{3-}-4e^{4U+2V_1}A^{9+}).
\end{equation}
From the form of ${Q_i}^j$, we can see that supersymmetry
corresponding to $\epsilon^{3,4}$ is broken for spherical and
hyperbolic $\Sigma_2$ since we cannot cancel the spin connections
along $\epsilon^{3,4}$. The $N=4$ supersymmetry is then broken to
$N=2$.
\\
\indent After using the condition \eqref{twist1} in the
${Q_{\hat{\phi}i}}^j$ components, the twist is achieved by imposing
the projection
\begin{equation}
\gamma^{\hat{\theta}\hat{\phi}}\epsilon^{\hat{i}}={\epsilon^{\hat{i}}}_{\hat{j}}\epsilon^{\hat{j}}\label{projector1}
\end{equation}
provided that we impose the following twist condition
\begin{equation}
2\sqrt{2}\kappa p^9=1\, .
\end{equation}
Indices $\hat{i},\hat{j}=1,2$ denote the Killing spinors
corresponding to the unbroken supersymmetry. From equation
\eqref{projector1}, the chirality condition on $\epsilon^{\hat{i}}$
implies that
\begin{equation}
\gamma^{\hat{0}\hat{r}}\epsilon^{\hat{i}}=-i{\epsilon^{\hat{i}}}_{\hat{j}}\epsilon^{\hat{j}}\, .\label{projector2}
\end{equation}
With these projections, we can write the $\delta
\psi^i_{\hat{\theta}}=0$ equation, which is the same as $\delta
\psi^i_{\hat{\phi}}$ equation, as
\begin{equation}
g'\gamma_{\hat{r}}\epsilon^{\hat{i}}-\frac{2}{3}A_1^{\hat{i}\hat{j}}\epsilon_{\hat{j}}+\frac{i}{2}(\mc{V}_\alpha)^*{\mc{V}_M}^{\hat{i}\hat{j}}(i\mc{H}^{M\alpha}_{\hat{0}\hat{r}}-\mc{H}^{M\alpha}_{\hat{\theta}\hat{\phi}}){\epsilon_{\hat{j}}}^{\hat{k}}\epsilon_{\hat{k}}=0
\end{equation}
where we have multiplied the resulting equation by $\gamma^{\hat{\theta}}$. We further impose the projector
\begin{equation}
\gamma_{\hat{r}}\epsilon^{\hat{i}}=e^{i\Lambda}\delta^{\hat{i}\hat{j}}\epsilon_{\hat{j}}\label{projector3}
\end{equation}
in which $e^{i\Lambda}$ is an $r$-dependent phase. By equation \eqref{projector2}, this projector implies
\begin{equation}
\gamma_{\hat{0}}\epsilon^{\hat{i}}=ie^{i\Lambda}\epsilon^{\hat{i}\hat{j}}\epsilon_{\hat{j}}\, .\label{projector4}
\end{equation}
It should be noted that there are only two independent projectors
given in \eqref{projector1} and \eqref{projector3}. Therefore, the
entire flows preserve $\frac{1}{4}$ supersymmetry. On the other
hand, the $AdS_2\times \Sigma_2$ vacua is $\frac{1}{2}$
supersymmetric since the $\gamma_{\hat{r}}$ projection is not needed
for constant scalars.
\\
\indent As a next step, we introduce the ``superpotential'' $\mc{W}$
and ``central charge'' $\mc{Z}$ defined respectively by the
eigenvalues of
\begin{equation}
\frac{2}{3}A^{\hat{i}\hat{j}}_1=\mc{W}_{\hat{i}}\delta^{\hat{i}\hat{j}}
\end{equation}
and
\begin{equation}
-\frac{i}{2}(\mc{V}_\alpha)^*{\mc{V}_M}^{\hat{i}\hat{j}}(i\mc{H}^{M\alpha}_{\hat{0}\hat{r}}
-\mc{H}^{M\alpha}_{\hat{\theta}\hat{\phi}}){\epsilon_{\hat{j}}}^{\hat{k}}=\mc{Z}_{\hat{i}}\delta^{\hat{i}\hat{k}}\,
.
\end{equation}
It should be emphasized that no summation is implied in the above two equations.
\\
\indent With all these, we obtain the BPS equation from $\delta\psi^{\hat{i}}_{\hat{\theta}}=0$ equation
\begin{equation}
e^{i\Lambda}g'-\mc{W}_i-\mc{Z}_i=0
\end{equation}
which gives
\begin{equation}
g'=|\mc{W}_i+\mc{Z}_i|\qquad \textrm{and}\qquad e^{i\Lambda}=\frac{\mc{W}_i+\mc{Z}_i}{|\mc{W}_i+\mc{Z}_i|}\, .
\end{equation}
\\
\indent Using all of these results, we find that equation $\delta\psi^{\hat{i}}_{\hat{0}}=0$ gives
\begin{equation}
e^{i\Lambda}(f'+i\hat{A}_te^{-f})-\mc{W}_i+\mc{Z}_i=0\, .
\end{equation}
Taking the real and imaginary parts leads to the following BPS equations
\begin{equation}
f'=\textrm{Re}[e^{-i\Lambda}(\mc{W}_i-\mc{Z}_i)]\label{f_prime}
\end{equation}
and
\begin{equation}
\hat{A}_t=e^f\textrm{Im}[e^{-i\Lambda}(\mc{W}_i-\mc{Z}_i)].\label{At_constraint}
\end{equation}
We now come to $\delta\psi^{\hat{i}}_{\hat{r}}=0$ equation which
gives the $r$-dependence of the Killing spinors. When combined with
$\delta\psi^{\hat{i}}_{\hat{0}}=0$ equation, this equation reads
\begin{equation}
2\epsilon^{\hat{i}\prime}-f'-i\hat{A}_te^{-f}\epsilon^{\hat{i}}=0
\end{equation}
which can be solved by
\begin{equation}
\epsilon^{\hat{i}}=e^{\frac{f}{2}+\frac{i}{2}\int \hat{A}_te^{-f}dr}\tilde{\epsilon}^{\hat{i}}.
\end{equation}
$\tilde{\epsilon}^{\hat{i}}$ are constant spinors satisfying the projections
\begin{equation}
\gamma_{\hat{r}}\tilde{\epsilon}^{\hat{i}}=\delta^{\hat{i}\hat{j}}\tilde{\epsilon}_{\hat{j}},\qquad
\gamma_{\hat{\theta}\hat{\phi}}\tilde{\epsilon}^{\hat{i}}={\epsilon^{\hat{i}}}_{\hat{j}}\tilde{\epsilon}^{\hat{j}}\,
.
\end{equation}
\indent Using the $\gamma_{\hat{r}}$ projector, we obtain the
following BPS equations from $\delta\chi^i$ and $\delta\lambda^i_a$
\begin{eqnarray}
-e^{i\Lambda}\epsilon^{\alpha\beta}\mc{V}_\alpha \mc{V}'_\beta\delta_{\hat{i}\hat{j}}-\frac{4i}{3}A^{\hat{j}\hat{i}}_2
+i\mc{V}_\alpha{\mc{V}_M}^{\hat{i}\hat{k}}\epsilon^{\hat{k}\hat{j}}(i\mc{H}^{M\alpha}_{\hat{0}\hat{r}}+\mc{H}^{M\alpha}_{\hat{\theta}\hat{\phi}})&=&0,\\
{\mc{V}_a}^M{\mc{V}_M}^{ij\prime}e^{-i\Lambda}+\frac{1}{4}\mc{V}_\alpha \mc{V}_{Ma}(\mc{H}^{M\alpha}_{\hat{0}\hat{r}}+i\mc{H}^{M\alpha}_{\hat{\theta}\hat{\phi}})\delta^i_{\hat{i}}\delta^j_{\hat{j}}\epsilon^{\hat{i}\hat{j}}
+{A_{2aj}}^i&=&0\, .
\end{eqnarray}
Note that there are four equations from $\delta\lambda^i_a$ for each
value of $a=1,2,3$, but $\delta\lambda^{i=3,4}_a$ do not get any
contribution from the gauge fields. However, the scalars appearing
in these equations cannot be consistently set to zero since
${A_{2aj}}^i$ is not diagonal in $ij$ indices.
\\
\indent It should be pointed out that the $N=3$ supersymmetric
$AdS_4$ vacuum corresponds to the Killing spinors $\epsilon^{2,3,4}$
while $\epsilon^1$ is the Killing spinor of the $N=1$ $AdS_4$
critical point. In the next section, we will look for possible
$AdS_2\times \Sigma_2$ solutions to the above BPS equations. As
mentioned before, in the twist given above, the supersymmetry
corresponding to $\epsilon^{3,4}$ is broken. Therefore, the
resulting $AdS_2\times \Sigma_2$ solutions will preserve only two
supercharges or half of the $N=1$ supersymmetry corresponding to
either $\epsilon^1$ or $\epsilon^2$. We will analyze these two cases
separately.

\section{Supersymmetric $AdS_2\times \Sigma_2$ solutions}\label{AdS2_solution}
In this section, we look for the $AdS_2\times \Sigma_2$ fixed points
of the above BPS flow equations with constant scalars. These
solutions should correspond to IR fixed points of the RG flows from
twisted compactifications of the dual $N=3$ and $N=1$ SCFTs in three
dimensions. They also describe near horizon geometries of BPS black
holes arising from M2-branes wrapped on $\Sigma_2$. Before giving
the solutions, we first discuss the conditions for obtaining the
$AdS_2$ fixed points.
\\
\indent At the $AdS_2\times \Sigma_2$ geometries, the scalars are
constant, and we can choose the gauge in which $A_t^{M\alpha}\sim
0$. Furthermore, the warped factor $g(r)$ is required to be
constant, $g'(r)=0$. Let $r_h$ be the position of the horizon, we
can summarized the conditions for $AdS_2\times \Sigma_2$ solutions
and their properties as follow
\begin{eqnarray}
f(r_h)=\frac{r_h}{L_{AdS_2}},\qquad e^{g(r_h)}&=&L_{\Sigma_2},\qquad \textrm{Im}[e^{-i\Lambda}(\mc{W}_i-\mc{Z}_i)]=0,\nonumber \\
|\mc{W}_i+\mc{Z}_i|&=&0,\qquad \frac{4}{3}A_2^{\hat{i}\hat{j}}=\mc{V}_\alpha {\mc{V}_M}^{\hat{i}\hat{k}}\epsilon^{\hat{k}\hat{j}}(i\mc{H}^{M\alpha}_{\hat{0}\hat{r}}+\mc{H}^{M\alpha}_{\hat{\theta}\hat{\phi}}),\nonumber \\
\frac{i}{4}\mc{V}_\alpha \mc{V}_{Ma}(-i\hat{H}^{M\alpha}_{\hat{0}\hat{r}}+\mc{H}^{M\alpha}_{\hat{\theta}\hat{\phi}})\epsilon^{\hat{i}\hat{j}}&=&-{A_{2a\hat{j}}}^{\hat{i}} \qquad {A_{2a\tilde{j}}}^{\hat{i}}=0,\quad \tilde{j}=3,4
\end{eqnarray}
where $L_{AdS_2}$ and $L_{\Sigma_2}$ are respectively the radii of
$AdS_2$ and $\Sigma_2$. These conditions can be viewed as attractor
equations for the scalars at the black hole horizon.

\subsection{Solutions in the $N=3$ case}
We begin with the $N=3$ case. The $AdS_2\times \Sigma_2$ solutions
will describe the fixed points of the RG flows from $N=3$ SCFTs dual
to the $N^{010}$ compactification of eleven-dimensional supergravity
to supersymmetric CFT$_1$'s dual to the $AdS_2\times \Sigma_2$
geometries. These flows are examples of the twisted
compactifications of the $N=3$ SCFT on $\Sigma_2$.
\\
\indent In this case, the superpotential and central charge are
given in term of the redefined scalars
$(\tilde{U}_1,\tilde{U}_2,\tilde{V})$ by
\begin{eqnarray}
\mc{W}_2&=&\frac{1}{2}e^{-\frac{1}{2}(4\tilde{U}_1+2\tilde{U}_2+\tilde{V})}\left[e^{2\tilde{U}_2}+4e^{\tilde{U}_1+\tilde{U}_2}-2e^{\tilde{U}_2+\tilde{V}}+4e^{\tilde{U}_1+\tilde{V}}
\right.\nonumber \\
& &  -3k+2iZ_2e^{\tilde{U}_2}+4iZ_2e^{\tilde{U}_1}-4iZ_1(e^{\tilde{U}_2}+e^{\tilde{V}}+iZ_2)\nonumber \\
& &\left.-2iZ_2e^{\tilde{V}}-Z_2^2+2\chi(2ie^{\tilde{U}_1}-ie^{\tilde{U}_2}+2Z_1+Z_2) \right],\\
\mc{Z}_2&=&\frac{1}{4}e^{-\frac{1}{2}(4g+2\tilde{U}_2+\tilde{V})}\left[2e_3e^{\tilde{U}_2}-\sqrt{2}ie_9e^{2\tilde{U}_2}+2ie_3\chi+2p^3\chi e^{\tilde{U}_2} \right.\nonumber\\
& &-\sqrt{2}ip^9\chi(e^{2\tilde{U}_2}+3k)-4\sqrt{2}\tilde{B}[e^{\tilde{U}_2}+e^{\tilde{V}}+i(\chi+Z_2)]\nonumber\\
& &+2ie_3Z_2+2\sqrt{2}e_9Z_2e^{\tilde{U}_2}+2ip^3\chi Z_2+2\sqrt{2}p^9\chi Z_2e^{\tilde{U}_2}\nonumber\\
& &+\sqrt{2}i(e_9+p^9\chi)Z_2^2+4iB(e^{2\tilde{U}_2}-2e^{\tilde{U}_2+\tilde{V}}-3k)\nonumber \\
& &+4B[2\chi (e^{\tilde{U}_2}+iZ_2)+Z_2(e^{\tilde{V}}-2e^{\tilde{U}_2}-iZ_2)]\nonumber \\
& &+e^{\tilde{V}}(2e_3-3\sqrt{2}p^9-\sqrt{2}p^9e^{2\tilde{U}_2}+2p^3Z_2+\sqrt{2}p^9Z_2^2)\nonumber \\
& &\left. -2ie^{\tilde{U}_2+\tilde{V}}(p^3+\sqrt{2}p^9Z_2)\right]
\end{eqnarray}
in which the subscript $2$ on $\mc{W}_2$ and $\mc{Z}_2$ refers to the
superpotential and central charge associated to the Killing spinor
$\epsilon^2$.
\\
\indent The BPS equations are given by
\begin{eqnarray}
f'&=&\textrm{Re}[e^{-i\Lambda}(\mc{W}_2-\mc{Z}_2)],\qquad e^{i\Lambda}=\frac{\mc{W}_2+\mc{Z}_2}{|\mc{W}_2+\mc{Z}_2|},\\
g'&=&|\mc{W}_2+\mc{Z}_2|,
\end{eqnarray}
\begin{eqnarray}
e^{i\Lambda}\tilde{V}'-ie^{-\tilde{V}+i\Lambda}\chi'&=&\frac{1}{2}
\left[e^{-\frac{\tilde{V}}{2}-\tilde{U}_2-2\tilde{U}_1}\left[2e^{\tilde{U}_2}+8e^{2\tilde{U}_1}
-6k+Z_2(8Z_1-2Z_2)
\right]\right.\nonumber \\
 &
&-e^{-2g-2\tilde{U}_1+\frac{\tilde{V}}{2}}\left[4e^{2g}+2e^{2\tilde{U}_1}(p^3+4B+\sqrt{2}p^9Z_2)\right]\nonumber
\\
& & \left.
+4\chi(2Z_1+Z_2)e^{-\frac{\tilde{V}}{2}-\tilde{U}_2-2\tilde{U}_1}+\sqrt{2}e_9e^{\tilde{U}_2-2g-\frac{\tilde{V}}{2}}\right]
 \nonumber \\
& &+\frac{1}{2}e^{-2g-\tilde{U}_2-\frac{\tilde{V}}{2}}\left[\sqrt{2}Z_2(4\tilde{B}-e_9Z_2)-2e_3(\chi+Z_2)+4\sqrt{2}\chi \tilde{B} \right.\nonumber\\
& &-4B(e^{2\tilde{U}_2}-3k+2\chi Z_2-Z_2^2)
+\sqrt{2}p^9\chi(e^{\tilde{U}_2}+3k)\nonumber \\
& &\left.-Z_2\chi(2p^3+\sqrt{2}p^9Z_2) \right]\nonumber \\
&
&-\frac{i}{2}e^{-\tilde{U}_2-\frac{\tilde{V}}{2}}\left[4e^{\tilde{U}_2-2\tilde{U}_1}(Z_2-2Z_1-\chi)
-4e^{\tilde{V}-2\tilde{U}_1}(2Z_1+Z_2) \right.\nonumber \\
&
&-2e^{\tilde{U}_2-2g}\left[Z_2(\sqrt{2}e_9-4B-2\sqrt{2}\tilde{B})+\chi
(p^3+4B+\sqrt{2}p^9Z_2)
\right]\nonumber \\
& &+
e^{\tilde{V}-2g}\left[2e_3-4\sqrt{2}\tilde{B}-\sqrt{2}p^9(3k+e^{2\tilde{U}_2})
-4\sqrt{2}\tilde{B}\right.\nonumber
\\
&
&\left.\left.+Z_2(2p^3+8B+\sqrt{2}p^9Z_2)\right]-2e^{\tilde{U}_2-2g}e_3\right],
\end{eqnarray}
\begin{eqnarray}
e^{-i\Lambda}\tilde{U}_2'+ie^{-\tilde{U}_2-i\Lambda}Z_2'&=&\frac{1}{2}e^{-2g-\tilde{U}_2
-2\tilde{U}_1-\frac{\tilde{V}}{2}}\left[2e^{2(g+\tilde{U}_2)}+\sqrt{2}ie_9e^{2(\tilde{U}_1+\tilde{U}_2)}
+6ke^{2g} \right.\nonumber \\
& &-2ie_3\chi e^{2\tilde{U}_1}+\sqrt{2}ip^9\chi e^{2(\tilde{U}_1+\tilde{U}_2)}
+3\sqrt{2}ikp^9\chi e^{2\tilde{U}_1}\nonumber \\
& &+8iZ_2e^{2g+\tilde{U}_1}-2ie_3Z_2e^{2\tilde{U_1}}-4\chi Z_2e^{2g}-2ip^3\chi Z_2e^{2\tilde{U}_1}\nonumber \\
&
&-8Z_1Z_2e^{2g}+2Z_2^2e^{2g}-\sqrt{2}ie_9Z_2^2e^{2\tilde{U}_1}-8\chi
Z_1 e^{2g}
\nonumber \\
&
&-4iBe^{2\tilde{U}_1}\left[e^{2\tilde{U}_2}-3k+Z_2(2\chi-Z_2-2ie^{\tilde{V}})\right]+8i\chi
e^{2g+\tilde{U}_1}\nonumber \\
&
&+4\sqrt{2}\tilde{B}e^{2\tilde{U}_1}(e^{\tilde{V}}+i\chi+iZ_2)-\sqrt{2}ip^9\chi
Z_2^2e^{2\tilde{U}_1}
\nonumber \\
& &
-4ie^{2g+\tilde{V}}(2Z_1+Z_2)-Z_2e^{2\tilde{U}_1+\tilde{V}}(2p^3+\sqrt{2}p^9Z_2)
\nonumber \\
&
&\left.-e^{\tilde{U}_1+\tilde{V}}\left[8e^{2g}+e^{\tilde{U}_1}\left[2e_3-\sqrt{2}p^9(e^{2\tilde{U}_2}+3k)\right]\right]
\right],
\end{eqnarray}
\begin{eqnarray}
e^{-i\Lambda}\tilde{U}_1'-ie^{-\tilde{U}_1-i\Lambda}Z_1'&=&e^{-\tilde{U}_2-2\tilde{U}_1-\frac{\tilde{V}}{2}}\left[2e^{\tilde{U}_2+\tilde{V}}-e^{2\tilde{U}_2}-2e^{\tilde{U}_1}(e^{\tilde{U}_2}+e^{\tilde{V}})+3k \right.\nonumber \\
& &-4iZ_1(e^{\tilde{U}_2}+e^{\tilde{V}}-iZ_2)+2i\chi (e^{\tilde{U}_1}-e^{\tilde{U}_2}+2iZ_1+iZ_2)\nonumber \\
& &\left.+2iZ_2(e^{\tilde{U}_2}+e^{\tilde{U}_1}-e^{\tilde{V}})+Z_2^2
\right]
\end{eqnarray}
where we have used the relation \eqref{twist1} to express $p^6$ in terms of $p^9$ and $e_3$.
\\
\indent To obtain the complete flow solutions, we have to solve
these equations together with the two-form equations
\eqref{2-form-eq1}, \eqref{2-form-eq2} and the equations for the
gauge fields \eqref{Ap1} to \eqref{Ap6} as well as the algebraic
constraint given by equation \eqref{At_constraint}. These equations
are very complicated even with the numerical technique not to
mention the analytic solutions. In what follow, we will present only
the $AdS_2\times\Sigma_2$ solutions and will not give the numerical
flow solutions which may be obtained by suitable boundary
conditions. In principle, the horizon is characterized by the values
of the scalars as functions of the electric and magnetic charges.
However, due to the complexity of the BPS equations, it is more
convenient to solve the horizon conditions for the charges in terms
of the scalar fields although inverting the solutions to express the
scalars in terms of the charges is desirable.
\\
\indent In the present case, although it is straightforward to solve
the above equations for $(B,\tilde{B},\chi, Z_1,p^9,p^3,e_3,e_9)$ in
terms of $(\tilde{U}_1, \tilde{U}_2, \tilde{V},Z_2)$, the resulting
expressions turn out to be cumbersome and not very illuminating.
Accordingly, we refrain from giving the general result here but
instead present some solutions with specific values of the
parameters. These are obtained from truncating the full result and
represent some examples of $AdS_2\times \Sigma_2$ geometries within the solution space.
\\
\indent Examples of $AdS_2\times \Sigma_2$ solutions are as
follow:
\begin{itemize}
\item We begin with a simple solution with vanishing pseudoscalars.
In the M-theory point of view, only scalars coming from the eleven-dimensional metric are turned on.
The solution is given by
\begin{eqnarray}
k&=&\frac{1}{5},\qquad \chi=Z_1=Z_2=0,\qquad e_9=0,\qquad \tilde{V}=\frac{1}{2}\ln \left[\frac{27}{5}\right],\nonumber \\
\tilde{U}_1&=&\frac{1}{2}\ln \left[\frac{27}{80}\right],\qquad
\tilde{U}_2=-\frac{1}{2}\ln \left[\frac{5}{3}\right],\qquad
\tilde{B}=\frac{1}{20}(5\sqrt{2}e_3-27p^9),\nonumber \\
g&=&\frac{1}{2}\ln\left[-\frac{81}{80}\sqrt{\frac{3}{10}}\kappa
p^9\right],\qquad B=-\frac{p^3}{4},\qquad
L_{AdS_2}=\frac{3^{\frac{9}{4}}}{32(5)^{\frac{3}{4}}}\, .
\end{eqnarray}
It is clearly seen that only the hyperbolic horizon ($\kappa=-1$) is possible otherwise
$g(r_h)$ will become complex. Therefore, we find that this is an $AdS_2\times H^2$ solution.
\item We next consider a solution with scalars and pseudoscalars turned on.
In the eleven-dimensional context, the solution involves scalar fields from both the metric
and the four-form field. This solution is characterized by
\begin{eqnarray}
k&=&1,\qquad Z_1=Z_2=\tilde{U}=0,\qquad \tilde{U}=\tilde{V}=\ln \left[\frac{12}{7}\right],\nonumber \\
p^3&=&\frac{41e_9+220p^9}{41\sqrt{2}},\qquad
B=-\frac{41e_9+136p^9}{164\sqrt{2}},\qquad
\tilde{B}=\frac{e_3}{2\sqrt{2}}-\frac{111}{41}p^9,\nonumber \\
\chi&=&-\frac{1}{7},\qquad
g=\frac{1}{2}\ln\left[-2^{\frac{5}{2}}\kappa p^9\sqrt{
\frac{21}{41}}\right],\qquad L_{AdS_2}=\frac{\sqrt{21}}{19}\, .
\end{eqnarray}
This solution is also $AdS_2\times H^2$.
\item As a final example, we consider a solution with more scalars turned on
and hence more general than the previous two solutions. This solution is given by
\begin{eqnarray}
Z_1&=&0,\qquad Z_2=-\frac{2\sqrt{k}}{7},\qquad \chi=-\frac{\sqrt{k}}{7},\qquad \tilde{U}_1=\tilde{U}_2
=\frac{1}{2}\ln k,  \nonumber \\
p^3&=&\frac{128,447k-104,895}{4,116\sqrt{2k}}p^9,\qquad e_9=\frac{32,723k-13,923}{4,116\sqrt{2k}}p^9,
 \nonumber \\
\tilde{B}&=&\frac{e_3}{2\sqrt{2}}+\frac{567-667k}{98}p^9,\qquad
g=\frac{1}{2}\ln \left[\frac{21(1-k)\sqrt{k}\kappa
p^9}{2\sqrt{2}}\right],\nonumber \\
\tilde{V}&=&\ln (2\sqrt{k}),\qquad
B=-25p^9\left[\frac{3,809k-2,961}{16,464\sqrt{2k}}\right], \qquad
L_{AdS_2}=\frac{k^{\frac{3}{4}}}{3\sqrt{2}}\, .\nonumber \\
& &
\end{eqnarray}
In this case, the flux parameter $k$ is not fixed, and there are two
types of solutions, $AdS_2\times S^2$ and $AdS_2\times H^2$,
depending on the value of $k$. For $k>1$, we have an $AdS_2\times
H^2$ solution with $\kappa=-1$ while the solution with $k<1$ is
$AdS_2\times S^2$ for which $\kappa=1$.
\end{itemize}

\subsection{Solutions in the $N=1$ case}
We now repeat a similar analysis for the $N=1$ case in which the
$N=1$ $AdS_4$ vacuum arises from the squashed $N^{010}$ manifold.
This critical point exists only for $k<0$, and the $AdS_2\times
\Sigma_2$ solutions would be IR fixed points of the twisted
compactifications of the dual $N=1$ SCFT. The superpotential and
central charge are given by
\begin{eqnarray}
\mc{W}_1&=&\frac{1}{2}e^{-\tilde{U}_2-2\tilde{U}_1-\frac{\tilde{V}}{2}}\left[e^{2\tilde{U}_2}
-4e^{\tilde{U}_1+\tilde{U}_2}-2e^{\tilde{V}}(e^{\tilde{U}_2}+2e^{\tilde{U}_1}) +4Z_1(Z_2-ie^{\tilde{U}_2}-ie^{\tilde{V}})\right. \nonumber \\
&
&\left.-3k+iZ_2(2e^{\tilde{U}_2}-4e^{\tilde{U}_1}-2e^{\tilde{V}}+iZ_2)+2\chi
(2Z_1+Z_2-ie^{\tilde{U}_2}-2ie^{\tilde{U}_1})\right],\nonumber \\
& &\\
\mc{Z}_1&=&\frac{1}{4}e^{-2g-\tilde{U}_2-\frac{\tilde{V}}{2}}\left[2e_3(e^{\tilde{U}_2}+i\chi)-\sqrt{2}ie_9e^{2\tilde{U}_2}+2p^3\chi e^{\tilde{U}_2}-3\sqrt{2}ikp^9\chi \right.\nonumber \\
& &-\sqrt{2}ip^9 \chi e^{2\tilde{U}_2}-4\sqrt{2}\tilde{B}(e^{\tilde{U}_2}+e^{\tilde{V}+i\chi+i Z_2})+2ie_3Z_2\nonumber \\
& &+2\sqrt{2}e_9Z_2e^{\tilde{U}_2}+2ip^3\chi Z_2+2\sqrt{2}p^9\chi Z_2+\sqrt{2}ie_9Z_2^2\nonumber \\
& &+\sqrt{2}ip^9\chi Z_2^2+4B[2\chi(e^{\tilde{U}_2}+iZ_2)+i(e^{2\tilde{U}_2}-2e^{\tilde{U}_2+\tilde{V}}-3k)]\nonumber \\
& &+4BZ_2(2e^{\tilde{V}}-2e^{\tilde{U}_2}-iZ_2)-2ie^{\tilde{U}_2+\tilde{V}}(p^3+\sqrt{2}p^9Z_2)\nonumber \\
& &\left. +e^{\tilde{V}}(2e_3-6\sqrt{2}p^9-\sqrt{2}p^9e^{2\tilde{U}_2}+2p^3Z_2+\sqrt{2}p^9Z_2^2)\right].
\end{eqnarray}
\indent The procedure is essentially the same, so we will just
present the result of $AdS_2\times \Sigma_2$ solutions and leave the
explicit form of the corresponding BPS equations to the appendix. In
this case, it turns out to be more difficult to find the solutions
in particular we have not found any solutions without the
pseudoscalars turned on. With some effort, we obtain the following
solutions:
\begin{itemize}
\item We begin with a simple solution in which all scalars have the same value as the $N=1$
supersymmetric $AdS_4$ vacuum
\begin{eqnarray}
k&=&-\frac{18}{11},\qquad Z_1=Z_2=\chi=0,\qquad \tilde{U}_1=\tilde{U}_2=\ln5-\frac{1}{2}\ln \left[\frac{55}{6}\right],\nonumber \\
\tilde{V}&=&-\frac{1}{2}\ln \left[\frac{55}{6}\right],\qquad B=-\frac{p^3}{4},\qquad \tilde{B}=\frac{e_3}{2\sqrt{2}},\qquad e_9=-\frac{14p^3}{5\sqrt{2}},\nonumber \\
g&=&\frac{1}{2}\ln \left[-\frac{10}{11}\sqrt{\frac{15}{11}}\kappa p^9\right],\qquad L_{AdS_2}=\frac{5^{\frac{5}{4}}}{2^{\frac{5}{4}}(3^{\frac{1}{4}})(11^{\frac{3}{4}})}\, .
\end{eqnarray}
The solution is of the $AdS_2\times H^2$ form.
\item We now give a more complicated solution
\begin{eqnarray}
k&=&-\frac{18}{11},\qquad Z_1=\chi=0,\qquad \tilde{U}_1=\tilde{V}=\ln\left[7\sqrt{-\frac{3k}{319}}\right],\nonumber \\
p^3&=&\sqrt{\frac{3}{638}}\left(\frac{p^9}{3,190\sqrt{-k}}\right)(567,365k-1,002,298),\nonumber \\
B&=&\sqrt{\frac{3}{638}}\left(\frac{p^9}{89,320\sqrt{-k}}\right)(13,987,355k-27,368,286),\nonumber \\
\tilde{B}&=&\frac{e_3}{2\sqrt{2}}+\frac{3p^9}{8,932}(63,162-32,267k),\qquad Z_2=-5\sqrt{-\frac{3k}{319}},
\nonumber \\
g&=&\ln\left[7\left(\frac{3}{638}\right)^{\frac{1}{4}}\sqrt{(k-2)\sqrt{-k}\kappa
p^9}\right],\nonumber \\
\tilde{U}_2&=&\frac{1}{2}\ln \left[-\frac{588k}{319}\right],\qquad
L_{AdS_2}=\frac{21(3^{\frac{1}{4}})}{11}\sqrt{\frac{7}{21}}\left(\frac{2}{29}\right)^{\frac{3}{4}}\,
.
\end{eqnarray}
This solution also gives $AdS_2\times H^2$ geometry. To show that
this leads to real solutions, we explicitly give one example of the
possible solutions
\begin{eqnarray}
Z_1&=&\chi=0,\qquad e_9=54.35,\qquad p^3=-11.56,\qquad \tilde{U}_1=\tilde{V}=-0.14,\nonumber \\
\tilde{U}_2&=&0.55,\qquad Z_2=-0.62,\qquad B=10.66,\qquad
\tilde{B}=-13.77+0.35e_3,\nonumber \\
 g&=&1.06\, .
\end{eqnarray}
\end{itemize}

\subsection{Uplift formulae}
We end this section by giving the uplift formulae for embedding the
previously found $AdS_2\times \Sigma_2$ solutions in eleven
dimensions. We first identify the vector and tensor fields in the
$N=4$ gauged supergravity and those obtained from the dimensional
reduction of eleven-dimensional supergravity on a tri-sasakian
manifold
\begin{eqnarray}
A^3_1&=&\sqrt{2}A^{9+},\qquad a^3_1=-\sqrt{2}A^{6+},\qquad c^3_1=A^{3+},\qquad \tilde{a}^3_1=-A^{3-},\nonumber \\
\tilde{c}^3_1&=&\sqrt{2}A^{9-},\qquad a^{12}_2=\sqrt{2}B^{18},\qquad c^3_2=B^{78}\, .
\end{eqnarray}
With this identification and the ansatz for the scalars and vector fields, the eleven-dimensional metric and the four-form field are given by
\begin{eqnarray}
ds^2_{11}&=&e^{-\frac{1}{3}(4\tilde{U}_1+2\tilde{U}_2+\tilde{V})}\left[-e^{2f}dt^2+dr^2+e^{2g}
(d\theta^2+F(\theta)^2d\phi^2)\right]\nonumber \\
& &+e^{\frac{1}{3}(2\tilde{U}_1+\tilde{U}_2-\tilde{V})}ds^2(B_{\textrm{QK}})+e^{\frac{2}{3}(\tilde{U}_1-\tilde{U}_2+\tilde{V})}
\left[(\eta^1)^2+(\eta^2)^2\right]\nonumber \\
& &+e^{\frac{2}{3}(\tilde{V}-2\tilde{U}_1-2\tilde{U}_2)}(\eta^3+\sqrt{2}\mc{A}^9_tdt
-\sqrt{2}p^9F'(\theta)d\phi)^2
\end{eqnarray}
and
\begin{eqnarray}
G_4&=&-\left[6ke^{-(4\tilde{U}_1+2\tilde{U}_2+\tilde{V})+f+2g}-\sqrt{2}B\mc{A}^{9\prime}_t-\sqrt{2}B'\mc{A}^9_t\right]F(\theta)dt\wedge dr\wedge d\theta \wedge d\phi\nonumber \\
& &+B'F(\theta)dr\wedge d\theta \wedge d\phi\wedge \eta^3+dZ_1\wedge (\eta^1\wedge J^1+\eta^2\wedge J^2)\nonumber \\
& &[\sqrt{2}(\tilde{\mc{A}}^{9\prime}_t+\chi \mc{A}^{9\prime}_t)dr\wedge dt+\sqrt{2}(e_9+\chi p^9-\sqrt{2}B)F(\theta)d\theta \wedge d\phi]\wedge \eta^1\wedge \eta^2\nonumber \\
& &[(\mc{A}^{3\prime}_t+\sqrt{2}Z_2\mc{A}^{9\prime}_t)dr\wedge dt+(p^3+\sqrt{2}p^9Z_2+2B)F(\theta)d\theta \wedge d\phi ]\wedge J^3\nonumber \\
& &+2(\chi+2Z_1)\eta^1\wedge \eta^2\wedge J^3+(dZ_2\wedge J^3+d\chi \wedge \eta^2\wedge \eta^2)\wedge (\eta^3-\sqrt{2}p^9F(\theta)d\phi)\nonumber \\
& &+2[(\mc{A}^3_t+\sqrt{2}\tilde{\mc{A}}^9_t)dt-(\sqrt{2}e_9+p^3)F(\theta)d\phi+4(2Z_1+Z_2)\textrm{vol}(B_{\textrm{QK}})
\nonumber \\
& &
+(\chi+Z_2)(\eta^3+\sqrt{2}\mc{A}^9_tdt-\sqrt{2}p^9F(\theta)d\phi)]\wedge(\eta^1\wedge
J^2-\eta^2\wedge J^1).\label{4_form_flux}
\end{eqnarray}

\section{Conclusions}\label{conclusions}
In this paper, we have found a number of $AdS_2\times \Sigma_2$
solutions in $N=4$ gauged supergravity with $SO(3)\ltimes
(\mathbf{T}^3,\hat{\mathbf{T}}^3)$ gauge group. The solutions can be
uplifted to M-theory since the $N=4$ gauged supergravity is a
consistent truncation of eleven-dimensional supergravity on a
tri-sasakian manifold. These $AdS_2\times \Sigma_2$ gemetries are
expected to arise from the near horizon limit of certain dyonic BPS
black holes which can be identified as holographic RG flows from
twisted compactifications of the dual $N=1,3$ SCFTs in the UV to
superconformal quantum mechanics corresponding to the $AdS_2$ geometry
in the IR. We have found that most of the solutions have hyperbolic
horizons, but some of them have spherical horizons depending on the
values of the four-form flux parameter. These solutions provide examples
of $AdS_2$ geometries from M-theory compactified on a tri-sasakian
manifold such as $N^{010}$ and are hopefully useful in the
holographic study of the $N=1,3$ Chern-Simons-Matter theories
in three dimensions. They should also be useful in the study of
black hole entropy along the line of recent results in \cite{Zaffaroni_BH_entropy,BH_entropy_benini,BH_entropy_Passias}. In
this aspect, the near horizon solutions given here are enough
although we have not constructed the full black hole solutions,
numerically. It would be interesting to compute the topologically
twisted index in the dual $N=1,3$ SCFTs and compare with the black
hole entropy computed from the area of the horizon $A\sim
L^2_{\Sigma_2}$.
\\
\indent The solutions found here might constitute only a small
number of all possible solutions due to the complexity of the
resulting BPS equations. It could be interesting to look for more
solutions or even to identify all possible black hole solutions to
this $N=4$ gauged supergravity similar to the analysis in $N=2$
gauged supergravity. For the case of $N^{010}$ manifold, there
exists an invariant two-form in addition to the universal forms on a
generic tri-sasakian manifold. This leads to an additional vector
multiplet, called Betti multiplet, in $N=4$ gauged supergravity.
This vector multiplet corresponds to a baryonic symmetry in the dual
SCFTs. Finding a reduction that includes the Betti multiplet and
$SU(3)$ non-singlet fields would be very useful in order to find
more interesting black hole and other holographic solutions. We
leave all these issues for future work. \vspace{1cm}
\\
{\large{\textbf{Acknowledgement}}} \\
The author would like to thank Davide Cassani for useful
correspondences and the Abdus Salam Centre for Theoretical Physics
for hospitality while most of this work has been done. This work is
supported by The Thailand Research Fund (TRF) under grant
RSA5980037.
\appendix
\section{Useful formulae}\label{appendix}
In this appendix, we collect some convention on t' Hooft matrices
and details on Yang-Mills equations and complicated BPS equations in
the $N=1$ case.

\subsection{`t Hooft matrices}
In converting $SO(6)$ vector indices $m,n$ to chiral spinor indices $i,j$, we use the following `t Hooft matrices
\begin{eqnarray}
G_1^{ij}&=&\left[
                     \begin{array}{cccc}
                       0 & 1 & 0 & 0 \\
                       -1 & 0 & 0 & 0 \\
                       0 & 0 & 0  & 1\\
                        0 & 0 & -1  & 0\\
                     \end{array}
                   \right],\,
G_2^{ij}=\left[
                     \begin{array}{cccc}
                       0 & 0 & 1 & 0\\
                       0 & 0 & 0 & -1\\
                       -1 & 0 & 0  & 0\\
                        0 & 1 & 0  & 0\\
                     \end{array}
                   \right],\,
G_3^{ij}=\left[
                     \begin{array}{cccc}
                       0 & 0 & 0 & 1\\
                       0 & 0 & 1 & 0\\
                       0 & -1 & 0  & 0\\
                        -1 & 0 & 0  & 0\\
                     \end{array}
                   \right],\nonumber \\
G_4^{ij}&=&\left[
                     \begin{array}{cccc}
                       0 & i & 0 & 0\\
                       -i & 0 & 0 & 0\\
                       0 & 0 & 0  & -i\\
                        0 & 0 & i  & 0\\
                     \end{array}
                   \right],\,
G_5^{ij}=\left[
                     \begin{array}{cccc}
                       0 & 0 & i & 0\\
                       0 & 0 & 0 & i\\
                       -i & 0 & 0  & 0\\
                        0 & -i & 0  & 0\\
                     \end{array}
                   \right],\,
G_6^{ij}=\left[
                     \begin{array}{cccc}
                       0 & 0 & 0 & i\\
                       0 & 0 & -i & 0\\
                       0 & i & 0  & 0\\
                        -i & 0 & 0  & 0\\
                     \end{array}
                   \right].\nonumber\\
                   & &
\end{eqnarray}

\subsection{Field equations of gauge fields}
In this section, we present the full equations of motion for the gauge fields $A^{M\alpha}$. Equation \eqref{YM1} gives
\begin{eqnarray}
-*\mc{D}\mc{H}^{3-}&=&e^{-4(2U+V_1)}\left[4Z_1(\Phi'+2Z_3Z_1')-4e^{4U+2V_1}Z_3'-8Z_1^2Z_3'\right]dr\nonumber \\
& &-8Z_1e^{-4(2U+V_1)}(2A^{3-}-\sqrt{2}A^{6+}-3\sqrt{2}kA^{9+}),\\
*\mc{D}\mc{H}^{6-}&=&3\sqrt{2}ke^{-4(2U+V_1)}(\Phi'+2Z_3Z_1'-2Z_1Z_3')dr\nonumber \\
& &+12ke^{-4(2U+V_1)}(3kA^{9+}+A^{6+}-\sqrt{2}A^{3-}),\\
*\mc{D}\mc{H}^{9-}&=&\sqrt{2}ke^{-4(2U+V_1)}(\Phi'+2Z_3Z_1'-2Z_1Z_3')dr\nonumber \\
& &+4e^{-4(2U+V_1)}(3kA^{9+}+A^{6+}-\sqrt{2}A^{3-})
\end{eqnarray}
while equation \eqref{YM2} leads to
\begin{eqnarray}
-*\mc{D}\mc{H}^{3+}&=&2e^{-4(2U+V_1)}(\Phi'+2Z_3Z_1'-2Z_1Z_3')dr\nonumber \\
& &+4e^{-4(2U+V_1)}(3kA^{9+}+A^{6+}-\sqrt{2}A^{3-}),\\
*\mc{D}\mc{H}^{6+}&=&4\sqrt{2}ke^{-4(2U+V_1)}[e^{4U+2V_1}Z_3'+2Z_1^2Z_3'-Z_1(\Phi'+2Z_3Z_1')]dr\nonumber \\
& &-16Z_1e^{-4(2U+V_1)}(3kA^{9+}+A^{6+}-\sqrt{2}A^{3-}),\\
*\mc{D}\mc{H}^{9+}&=&0
\end{eqnarray}
For equations obtained from \eqref{YM3}, it is more convenient to express them in the following combinations
\begin{eqnarray}
\mc{H}^{9-}&=&e^{-4U+2V_1-V_2}(Z_2^2*\mc{H}^{9+}+*\mc{H}^{6+}+\sqrt{2}Z_2*\mc{H}^{3+})\nonumber \\
& &-\chi\mc{H}^{9+},\\
Z_2^2\mc{H}^{9-}+\mc{H}^{6-}+\sqrt{2}Z_2\mc{H}^{3-}&=&e^{4U+2V_1+3V_2}*\mc{H}^{9+}\nonumber \\
& &-\chi (Z_2^2\mc{H}^{9+}+\mc{H}^{6+}+\sqrt{2}Z_2\mc{H}^{3+}),\\
\sqrt{2}Z_2\mc{H}^{9-}+\mc{H}^{3-}&=&-\chi(\sqrt{2}Z_2\mc{H}^{9+}+\mc{H}^{3+})\nonumber
\\
& &-e^{2V_1+V_2}(\sqrt{2}Z_2*\mc{H}^{9+}+*\mc{H}^{3+}).
\end{eqnarray}

\subsection{BPS equations for the $N=1$ case}
In this section, we collect all the relevant BPS equations in the
$N=1$ case. These are given by
\begin{eqnarray}
e^{-i\Lambda }\tilde{U}'_1+ie^{-\tilde{U}_1-i\Lambda}Z_1'&=&e^{-\tilde{U}_2-2\tilde{U}_1-\frac{\tilde{V}}{2}}\left[2e^{\tilde{U}_1+\tilde{U}_2} -e^{2\tilde{U}_2}+2e^{\tilde{V}}(e^{\tilde{U}_2}+e^{\tilde{U}_1})+3k\right.\nonumber \\
& &-4iZ_1(e^{\tilde{U}_2}+e^{\tilde{V}}-iZ_2)-2i\chi (e^{\tilde{U}_2}+e^{\tilde{U}_1}-2iZ_1-iZ_2)\nonumber \\
& &\left.+Z_2[Z_2-2i(e^{\tilde{V}}+e^{\tilde{U}_1}-e^{\tilde{U}_2})] \right],
\end{eqnarray}
\begin{eqnarray}
e^{-i\Lambda}\tilde{U}'_2+ie^{-\tilde{U}_2-i\Lambda}Z_2'&=&\frac{1}{2}e^{-2g-\tilde{U}_2-2\tilde{U}_1-\frac{\tilde{V}}{2}}\left[2e^{2(g+\tilde{U}_2)}+\sqrt{2}ie_9e^{2(\tilde{U}_1+\tilde{U}_2)}+6ke^{2g} \right.\nonumber \\
& &-2ie_3\chi e^{2\tilde{U}_1}+\sqrt{2}ip^9\chi e^{2(\tilde{U}_1+\tilde{U}_2)}
+3\sqrt{2}ikp^9\chi e^{2\tilde{U}_1}\nonumber \\
& &-8iZ_2e^{2g+\tilde{U}_1}-2ie_3Z_2e^{2\tilde{U}_1}-4\chi Z_2e^{2g}-2ip^3\chi Z_2e^{2\tilde{U}_1}\nonumber \\
& &-8Z_1Z_2e^{2g}+2Z_2^2e^{2g}-\sqrt{2}ie_9Z_2^2e^{2\tilde{U}_1}-\sqrt{2}ip^9\chi Z_2^2e^{2\tilde{U}_1}\nonumber \\
& &-4iBe^{2\tilde{U}_1}[e^{2\tilde{U}_2}-3k+Z_2(2\chi-Z_2-2ie^{\tilde{V}})]-8i\chi e^{2g+\tilde{U}_1}\nonumber \\
& &-4ie^{2g+\tilde{V}}(2Z_1+Z_2)+4\sqrt{2}\tilde{B}e^{2\tilde{U}_1}[e^{\tilde{V}}+i(\chi+Z_2)]\nonumber \\
& &+e^{\tilde{U}_1+\tilde{V}}[8e^{2g}+e^{\tilde{U}_1}(\sqrt{2}p^9(e^{2\tilde{U}_2}+3k)-2e_3)]\nonumber \\
& &\left.-8\chi Z_1e^{2g}
-Z_2e^{2\tilde{U}_1+\tilde{V}}(2p^3+\sqrt{2}p^9Z_2) \right],
\end{eqnarray}
\begin{eqnarray}
e^{i\Lambda}\tilde{V}'-ie^{-\tilde{V}+i\Lambda}\chi'&=&\frac{1}{2}e^{-2g-\tilde{U}_2-2\tilde{U}_1-\frac{\tilde{V}}{2}}\left[2e^{2(g+\tilde{U}_2)}-8e^{2g+\tilde{U}_2+\tilde{U}_1}+2ie_3e^{\tilde{U}_2+2\tilde{U}_1} \right.\nonumber \\
& &+\sqrt{2}e_9e^{2(\tilde{U}_1+\tilde{U}_2)}-4i\chi e^{2g+\tilde{U}_2}-8i\chi e^{2g+\tilde{U}_1}-2e_3\chi e^{2\tilde{U}_1}\nonumber \\
& &+2ip^3\chi e^{\tilde{U}_2+2\tilde{U}_1}+\sqrt{2}p^9\chi e^{2(\tilde{U}_1+\tilde{U}_2)}
+3\sqrt{2}kp^9\chi e^{2\tilde{U}_1}\nonumber\\
& &+4 e^{2g}(2\chi Z_1+iZ_2e^{\tilde{U}_2})-2Z_2e^{\tilde{U}_1}(4ie^{2g}+e_3e^{\tilde{U}_1})-2Z_2^2e^{2g}\nonumber \\
& &+2\sqrt{2}ie_9Z_2e^{\tilde{U}_2+2\tilde{U}_1}+4\chi Z_2e^{2g}-2p^3\chi Z_2e^{2\tilde{U}_1}-6ke^{2g}
\nonumber \\
& &+2\sqrt{2}ip^9\chi Z_2e^{\tilde{U}_2+2\tilde{U}_1}+8Z_1Z_2e^{2g}-\sqrt{2}e_9Z_2^2e^{2\tilde{U}_1}
\nonumber \\
& &+4\sqrt{2}\tilde{B}e^{2\tilde{U}_1}[Z_2+\chi-i(e^{\tilde{U}_2}-e^{\tilde{V}})]-4Be^{2\tilde{U}_1}(e^{2\tilde{U}_2+\tilde{V}}-3k)\nonumber \\
& &-4Be^{2\tilde{U}_1}[2iZ_2(e^{\tilde{U}_2}+e^{\tilde{V}})-Z_2^2+2\chi (Z_2-ie^{\tilde{U}_2})]\nonumber \\
&
&+ie^{2\tilde{U}_1+\tilde{V}}\left[6\sqrt{2}p^9-2e_3+\sqrt{2}p^9-2p^3Z_2-\sqrt{2}p^9Z_2^2\right.
\nonumber\\
& & \left.
+2ie^{\tilde{U}_2}(p^3+\sqrt{2}p^9Z_2)\right]+4e^{2g+\tilde{V}}
[e^{\tilde{U}_2}+i(2Z_1+Z_2)]\nonumber
\\
&
&\left.+8e^{2g+\tilde{U}_1+\tilde{V}}-8iZ_1e^{2g+\tilde{U}_2}-\sqrt{2}p^9\chi
Z_2^2e^{2\tilde{U}_1}\right]
\end{eqnarray}
where
\begin{equation}
e^{i\Lambda}=\frac{\mc{W}_1+\mc{Z}_1}{|\mc{W}_1+\mc{Z}_1|}\, .
\end{equation}
These equations need to be solved together with the following equations
\begin{equation}
f'=\textrm{Re}[e^{-i\Lambda}(\mc{W}_1-\mc{Z}_1)],\qquad g'=|\mc{W}_1+\mc{Z}_1|  ,\qquad \hat{A}_t=e^f\textrm{Im}[e^{-i\Lambda}(\mc{W}_1-\mc{Z}_1)]
\end{equation}
and the two-form equations \eqref{2-form-eq1} and \eqref{2-form-eq2}.


\end{document}